\documentclass{aa}  

\usepackage{graphicx}
\usepackage{lipsum} 
\usepackage{txfonts}
\usepackage{amssymb}
\usepackage{amsmath}
\usepackage{float}
\usepackage[]{natbib}
\usepackage{multirow}
\usepackage{xcolor}

\begin{document} 

   \title{Investigating the presence of two belts in the HD\,15115 system
\thanks{Based on data collected at the European Southern Observatory, Chile under programs 098.C-0686 and 096.C-0640}}
%   \subtitle{ \thanks{B\textbf{•}} }

   \author{N.~Engler\inst{\ref{instch1}} 
   \and A.~Boccaletti\inst{\ref{instf4}} 
   \and H.M.~Schmid\inst{\ref{instch1}} 
   \and J.~Milli\inst{\ref{insteso2}}  
   \and J.-C.~Augereau\inst{\ref{instf1}}  
   \and J.~Mazoyer\inst{\ref{instus4},\ref{fellow}} 
   \and A.-L.~Maire\inst{\ref{instd1}} 
   \and T.~Henning\inst{\ref{instd1}} 
   \and H.~Avenhaus\inst{\ref{instd1}} 
   \and P.~Baudoz\inst{\ref{instf4}} 
   \and M.~Feldt\inst{\ref{instd1}}  
   \and R.~Galicher\inst{\ref{instf4}}  
   \and S.~Hinkley\inst{\ref{instuk1}}
   \and A.-M.~Lagrange\inst{\ref{instf8}}                
   \and D.~Mawet\inst{\ref{instus2}} 
   \and J.~Olofsson\inst{\ref{instchi1},\ref{instd1},\ref{instchi2}}
   \and E.~Pantin\inst{\ref{instf6}}
   \and C.~Perrot\inst{\ref{instf4}} 
   \and K.~Stapelfeldt\inst{\ref{instus4}} 
 }

\institute{
ETH Zurich, Institute for Particle Physics and Astrophysics, Wolfgang-Pauli-Strasse 27, 
CH-8093 Zurich, Switzerland, \email{englern@phys.ethz.ch}\label{instch1}
\and
LESIA, Observatoire de Paris, PSL Research Univ., CNRS, Univ. Paris Diderot, Sorbonne Paris Cité, UPMC Paris 6, Sorbonne Univ., 5 place Jules Janssen, 92195 Meudon, France\label{instf4}
\and
European Southern Observatory, Alonso de Cordova 3107, Casilla
19001 Vitacura, Santiago 19, Chile\label{insteso2}
\and
Universit\'{e} Grenoble Alpes, CNRS, IPAG, 38000 Grenoble, France\label{instf1}
\and
Jet Propulsion Laboratory, California Institute of Technology, 4800 Oak Grove Drive, Pasadena, CA 91109, USA\label{instus4}
%\and
%Space Telescope Science Institute, US\label{instus1}
\and
Max-Planck-Institut f\"{u}r Astronomie, K\"{o}nigstuhl 17, 69117
Heidelberg, Germany\label{instd1}
\and
Astrophysics Group, School of Physics, University of Exeter, UK\label{instuk1}
\and
Laboratoire d’astrophysique de Grenoble, Observatoire
de Grenoble, 38000 Grenoble, France\label{instf8}
\and
Instituto de Física y Astronomía, Facultad de Ciencias, Universidad de Valparaíso, Av. Gran Bretaña 1111, Playa Ancha, Valparaíso, Chile\label{instchi1}
\and
N\'ucleo Milenio Formaci\'on Planetaria - NPF, Universidad de Valpara\'iso, Av. Gran Breta\~na 1111, Valpara\'iso, Chile \label{instchi2}
\and
California Institute of Technology, Astronomy Option, USA\label{instus2}
\and
Centre d’Etudes de Saclay, France\label{instf6}
\and
Sagan NHFP Fellow\label{fellow}
}

   \date{Received ...; accepted ...}

\abstract
%context
{High-contrast instruments like SPHERE enable spatial resolution of young planetary systems and allow us to study the connection between planets and the dust contained in debris disks by the gravitational influence a planet can have on its environment.}
% aim
{We present new observations of the edge-on debris disk around HD\,15115 (F star at 48.2 pc) obtained in the near-IR. We search for observational evidence for a second inner planetesimal ring in the system. }
% method
{We obtained total intensity and polarimetric data in the broad bands J and H and processed the data with differential imaging techniques achieving an angular resolution of about 40~mas. A grid of models describing the spatial distribution of the grains in the disk is generated to constrain the geometric parameters of the disk
and to explore the presence of a second belt. We perform a photometric analysis 
of the data and compare disk brightness in two bands in scattered and in polarized light.
}
% results
{We observe an axisymmetric planetesimal belt with an inclination of $85.8^{\circ} \pm 0.7^{\circ}$ and position angle of $\sim$278.9$^{\circ} \pm 0.1^{\circ}$. The photometric analysis shows that the west side is $\sim$2.5 times brighter in total intensity than the east side in both bands, while for polarized light in the J band this ratio is only 1.25. We also find that the J - H color of the disk appears to be red for the radial separations $r\lesssim2''$ and is getting bluer for the larger separations. The maximum polarization fraction is 15--20\% at $r\sim$2.5$''$. The polarized intensity image shows some structural features inside the belt which can be interpreted as an additional inner belt.
}
% Conclusions
{The apparent change of disk color from red to blue with an increasing radial separation from the star could be explained by the decreasing average grain size with distance.\\
The presence of an inner belt slightly inclined with respect to the main planetesimal belt is suspected from the data but the analysis and modeling presented here cannot establish a firm conclusion due to the faintness of the disk and its high inclination.
}

%  \abstract

   \keywords{planetary systems: debris disks --
                stars: individual object: HIP 11360, HD 15115 --
                Techniques: high angular resolution, polarimetric
               }

\authorrunning{Engler et al.}

\titlerunning{HD\,15115
}

\maketitle
%
%________________________________________________________________

\section{Introduction} \label{s_Introduction}
\subsection{ Debris disks with multiple belts}

Due to the discovery of hundreds of debris disks by the Infrared Astronomical Satellite \citep[IRAS; e.g.][]{Aumann1985} and thousands of exoplanets by the current large missions like Kepler \citep[e.g.][]{Borucki2010} we know that planetary systems are not rare in the solar neighbourhood.

There is a growing evidence from the infrared and submillimeter surveys of the last few years that debris systems with multiple planetesimal belts around young main sequence stars are common as well \citep{Carpenter2009, Moor2011, Morales2013, Chen2014, Jang-Condell2015}. Their architectures could be similar to that of the solar system harbouring two prominent rings of debris: the main asteroid belt and the Edgeworth-Kuiper belt. 

Several hypotheses were recently developed to explain the formation of such systems. The most favourable of them are the disk sculpting by planets into distinct planetesimal belts \citep[e.g.][and references therein]{Dipierro2016, Geiler2017} and the interaction of dust grains with the gas in disk \citep[e.g.][]{Lyra2013}. The former mechanism assumes the presence of planets formed in the protoplanetary disk. In analogy to the solar system, the planets carve ring-like structures in the disk removing the gas and debris from their immediate vicinities, the so-called chaotic zones. They create a wide gap depleted of debris material which separates the regions of the disk still containing dust and planetesimals. Evidence for this hypothesis could be  provided by the planetary systems HR 8799 and HD 95086. In HR~8799, four planets \citep{Marois2010} reside in the hole between the warm and cold dust belts \citep{Su2009}. 
A similar disk configuration with a warm and a cold component \citep{Su2015} and a planet located in between them is detected in the HD 95086 system \citep{Rameau2013}.

Multiple ring structures can also originate from local instabilities induced by the gas in dusty disks. Using the hydrodynamical simulation of the gas-disk interaction, \citet{Lyra2013} showed that the photoelectric heating of the gas by locally concentrated dust grains can create a pressure maximum resulting in a further dust concentration at this location. 
However, in order for this mechanism to work, a strong coupling of dust to the gas is required and therefore a disk dust-to-gas ratio lower than unity. This condition might be fulfilled in the so-called transitional disks such as HD 141569A \citep{Thi2014} or in the gas-rich debris disks such as HIP\,73145 \citep{Moor2017} in which multiple ring structures were observed \citep{Perrot2016, Konishi2016, Feldt2017}.

The existence of double planetesimal belts is often inferred by analysing the stellar spectral energy distribution (SED). The SEDs of such systems show an infrared (IR) excess on top of the stellar photosphere which can be well fitted by a combination of two Planck's laws with different temperatures. This implies that at least two spatially separated populations of dust exist in the system \citep[e.g.][]{Morales2011, Ballering2013}, a so-called hot/warm component located closer than 10\,au to the host star and a cold component usually between 70 and 150\,au away.

Generally, the SED modeling can provide neither the actual number of belts in the system nor an accurate estimate of their location and spatial extent due to degeneracy between orbital distance and grain properties. Multiple cold belts, for instance, do not necessarily show up as distinct features in the SED. The actual distribution of the dust can only be revealed with direct imaging allowing high-spatial resolution and high contrast \citep[e.g.][]{Matthews2017}. Usually, the innermost component is inaccessible for direct imaging because it is faint and located too close to the star. To date, only a few examples of debris disks with multiple belts have been spatially resolved in scattered light; for example HD 92945 \citep{Golimowski2011}, HIP\,73145 \citep{Feldt2017}, HIP\,67497  \citep{Bonnefoy2017} and HD\,141569A \citep{Perrot2016}. In all these cases, the positions of the belts measured by direct imaging are different from those inferred from the SED modeling. 

For a successful detection of multiple belts, a low inclination of the system is an advantage. All the above-mentioned objects have an inclination $i$ lower than $80^\circ$. Obviously, it is much more difficult to identify more than one belt in edge-on disks. In this work, we investigate the presence of two belts of planetesimals in the well-known edge-on disk around HD\,15115.

\subsection{Previous observations of HD 15115} 
HD\,15115 (HIP\,11360) is a nearby zero-age main sequence F4IV star \citep{Harlan1974}. According to the Gaia DR1 catalogue \citep{GaiaCollaboration2016}, HD\,15115 has a parallax angle of $20.76\pm 0.41$\,mas corresponding to a distance of $48.16\pm 0.95$\,pc which we apply in this work. This value is slightly different from the former determination of $45.2\pm 1.3$~pc \citep{vanLeeuwen2007}. The age of the star is very uncertain with a range from 25 Myr, as a member of the $\beta$ Pictoris moving group \citep[BPMG;][]{Mamajek2014}, although the BPMG membership of HD\,15115 has been questioned \citep{Debes2008}, to 100--500 Myr, based on different age indicators \citep{Debes2008}. 

The IR excess indicating thermal emission from circumstellar dust was first detected with IRAS \citep{Silverstone2000}. Using the \textit{Hubble Space Telescope} (\textit{HST}) Advanced Camera for Surveys (ACS) and the Keck near-IR camera NIRC2, \citet{Kalas2007} subsequently discovered a highly asymmetrical debris disk in the optical and in the near-IR (H band). The disk was found to be nearly edge-on extending along the east-west direction with a west side detected out to a stellocentric radius of $\approx 580$\,au ($\approx12''$) and the east side to at least 340\,au ($\approx 7''$). 

The complexity of the inner disk structure was noted by \citet{Debes2008} who imaged HD\,15115 with the \textit{HST} Near-Infrared and Multi-Object Spectrometer (NICMOS) at 1.1 $\mu$m and claimed a possible ``warp'' on the western side of the disk at around $1''$. They also established a wavelength dependence of the brightness asymmetries. \citet{Kalas2007} measured an overall blue V - H color for the west side, which together with the shape of the disk gave it its nickname: ``the blue needle''. \citet{Debes2008} detected a relatively intense variation of disk colors as a function of distance from the central star: at $2''$ they measured a neutral color up to 1.1~$\mu$m and a blue spectrum for $\lambda > 1.1$~$\mu$m, while at $1''$ two disk extensions show a red F110W - H color. This suggests the presence of several populations of grains with various properties, such as grain size or composition, at different separations. The red color of the inner disk could be due to the strong absorption features of olivine at 1 $\mu$m \citep{Debes2008}. \cite{Rodigas2012} found, on the contrary, that the disk possesses mostly a spectrally neutral K$_S$ - L$'$ color (hence, ``the grey needle''). 

\begin{table*} 
      \caption[]{Summary of observations.}
         \label{Settings}
         \begin{tabular}{cccccccccc}
            \hline 
            \hline
            Observation & Instrument &  Filter & Field & \multicolumn{2}{c}{Integration Time}& & \multicolumn{3}{c}{Observing Conditions} \\\cline{5-6} \cline{8-10}
            date & mode &  & rotation [$^{\circ}$] & DIT [s]&Total [min]& &Airmass &Seeing ["] & Coh. time [ms]  \\
            \hline
            \hline
            \noalign{\smallskip}
            2015-10-29 & IRDIS CI &  BB\_H & 23.2 & 16 & 51.2 &&$1.18\pm 0.02$&$0.92\pm 0.09$&$1.7\pm 0.2$\\
            2015-11-19 & IRDIS CI &  BB\_J & 23.5 & 8 & 51.2 &&$1.18\pm 0.02$&$0.78\pm 0.10$&$1.6\pm 0.3$\\
            2016-10-03 & IRDIS DPI &BB\_J &stabilized& 32 & 85.3 &&$1.20\pm 0.03$&$0.87\pm 0.18$&$2.2\pm 0.4$\\

            \hline
            \hline
            \noalign{\smallskip}
         \end{tabular}
         
\end{table*}
   
Analysing the data obtained with PISCES on the Large Binocular Telescope (LBT) and LBTI/LMIRcam, these latter authors showed that the east-west asymmetry is still present in the K$_S$ band but the disk appears to be almost symmetrical in the L$'$ band. This symmetry was confirmed by \cite{Mazoyer2014} who detected a ring-shaped structure with a radius of $R\approx1.99''$ and a position angle (PA) of $98.8^\circ$ in H and K$_S$ band images from the Gemini Near-Infrared Coronagraphic Imager (NICI) archival data sets. They also found that the disk has an inclination of $\sim$86$^\circ$, shows no noticeable center offset with respect to the star and probably has an inner cavity. This work provided the first unbiased view of the disk structure though with some discrepancies with the results of recent observations of HD\,15115 with $HST$ \citep{Schneider2014}. \cite{Schneider2014} confirm the bow-like structure measured by \cite{Rodigas2012} and \cite{Mazoyer2014} but obtain a different inclination of $\sim80^\circ$ for an inner ``half-ring'' while they estimate that the extended disk is edge-on \citep[][see their Table 5]{Schneider2014}. The authors also note a bifurcation of the scattered light away from the major axis on the east disk side and a significantly larger vertical extension of the disk on the west side compared to the east. 

These puzzles could not be solved with the 1.3 mm observations using the Submillimeter Array \citep[SMA;][]{MacGregor2015}. The resolved millimeter emission showed a rather symmetric distribution of large grains in the disk with a tentative detection of the asymmetric extension towards the west. \citet{MacGregor2015} fitted these data with a model of a broad planetesimal belt extending from $R_{\rm in}=43$ to $R_{\rm out}=113$\,au.

Such a broad debris belt could be a combination of two or even more narrower planetesimal belts remaining unresolved in the SMA data. Previous studies have already classified the HD 15115 disk as a double belt system based on the SED modeling. 
\cite{Moor2011} used the disk fluxes at $\lambda= 24$ $\mu$m, $\lambda= 70$ $\mu$m, and $\lambda= 160$ $\mu$m (Table~\ref{t_SED}) as well as the {\it Spitzer} Infrared Spectrograph (IRS) 5-35 $\mu$m spectrum for their fit and found two belts with radii  $R_{\rm warm \, dust} = 4$ au and $R_{\rm cold \, dust} = 42$ au. \cite{Chen2014} derived $R_{\rm warm \, dust} = 5$ au and $R_{\rm cold \, dust} = 175$ au based on the fits to the same {\it Spitzer} IRS spectrum and the IR excess at $\lambda= 13$ $\mu$m, $\lambda= 24$ $\mu$m, and $\lambda= 31$ $\mu$m.

In this paper, we present new observations of HD\,15115 in the broadbands J and H
carried out with SPHERE (Spectro-Polarimetric High-contrast Exoplanet REsearch;
\cite{Beuzit2008}) at the Very Large Telescope (VLT) in Chile.
The paper is organized as follows. In Sect.\,\ref{Observations} we describe the observations 
and present the data. Section \ref{data} is dedicated to the methods of data reduction and Sect.~\ref{Data Analysis} to the data analysis. The modeling of the disk geometry using both the total intensity and the polarimetric data is presented in Sect. \ref{s_modeling}. Section~\ref{s_Photometry} describes the measurement of stellar and disk fluxes and compares the results from different datasets. In Sect.~\ref{Discussion}, we discuss our data analysis and possible indications for multiple belts in the HD\,15115 system from earlier observations and finally summarize our results in Sect.~\ref{s_Summary}.

\section{Observations} \label{Observations}
The HD\,15115 observations (098.C-0686 and 096.C-0640) were performed with the Infra-Red Dual-beam Imager and Spectrograph \citep[IRDIS,][]{Dohlen2006}, the near-infrared imaging sub-instrument of SPHERE. 
We used the pupil-stabilized mode of IRDIS in order to allow for angular differential imaging \citep[ADI;][]{Marois2006} while the field of view rotates. The observations of HD\,15115 with the broad-band filter BB\_H ($\lambda_c=1626$\,nm, $\Delta\lambda= 291$\,nm, hereafter H band) in both channels of the instrument were conducted on Oct 29, 2015, and with the broad-band filter BB\_J ($\lambda_c=1258$\,nm, $\Delta\lambda= 197$\,nm, hereafter J band) on Nov 19, 2015. To block the stellar light we used an apodized Lyot coronagraph (N\_ALC\_YJH\_S) including a focal mask (ALC2) with a diameter of 185\,mas \citep[inner working angle of 95\,mas,][]{Boccaletti2018}. The detector integration time (DIT) and the total exposure time together with the observing conditions are summarized in Table \ref{Settings}. To measure the stellar flux, several non-coronagraphic frames were recorded before and after each observation with DIT~$\approx$~0.84 s in H band and DIT\,= 2.00 s in J band using a neutral density filter ND2.0 with a $\sim$10$^{-3} - 10^{-2}$ transmissivity to prevent the saturation of the detector. 

At the beginning of the science observation a ``center frame'' was taken using the deformable mirror waffle mode \citep{Langlois2013}. This frame provides the accurate measurement of the star position when using the coronagraph. The stability of the star position during the observation is ensured by the Differential Tip-Tilt Sensor \citep[DTTS,][]{Baudoz2010}. 

For the polarimetric measurements on Oct 3, 2016, we used the dual-polarization imaging mode (DPI) of IRDIS \citep{Langlois2010}, for which the field of view is stabilized. The data were taken in the J band with the same apodized Lyot coronagraph as in the imaging mode. We performed ten $QU$ cycles consisting of consecutive measurements of Stokes parameters $Q, -Q, U, -U$  setting DIT to 32s per frame. The total exposure time of the polarimetric observations was 5120s. As for the first observations, the stellar flux in the J band is obtained with an off-axis image of the star using a neutral density filter. 

A raw frame with imaging or polarimetric data recorded with IRDIS has a format of 2048$\times$1024 pixels. In Classical Imaging (CI), raw frames contain two images side-by-side (left and right channel) simultaneously acquired with a broadband filter in the common path. In DPI mode, the left and right images contain the intensities of orthogonal linear polarizations, also taken simultaneously with a common broadband filter. The common optical path of both channels ensures that the two beams have similar optical aberrations allowing an efficient suppression of speckles within the AO (Adaptive Optics) control radius.

The pixel scale of the IRDIS detector is 12.25 mas and the field of view (FOV) is $11'' \times 11''$.

\begin{figure*} 
   \centering

  \includegraphics[width=17.5cm]{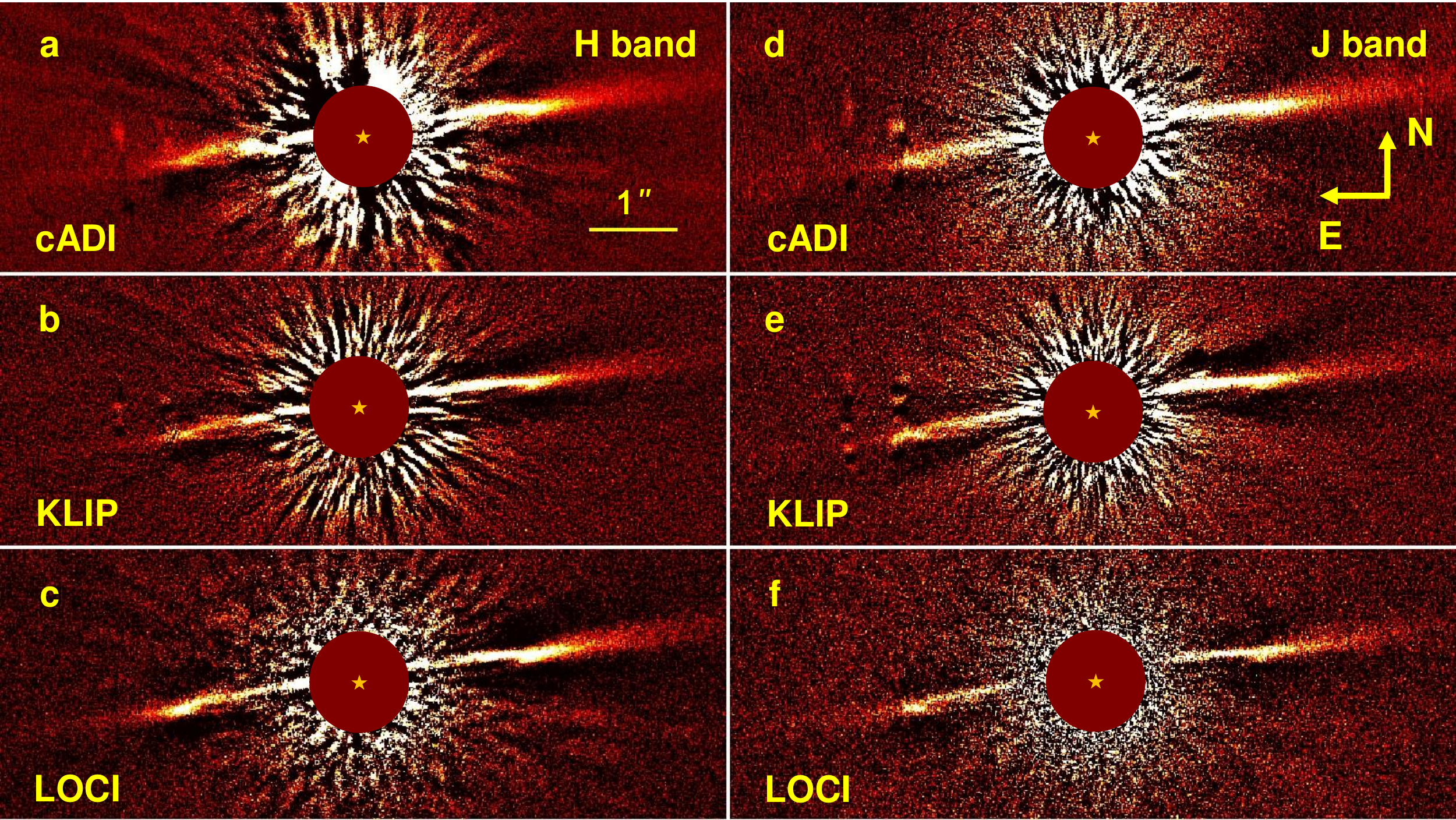}  
   \caption{Total intensity images of the HD\,15115 debris disk obtained with different data-reduction techniques in the H band ({\it left column}) and in the J band ({\it right column}). 
   The number of modes in KLIP reductions is 10. The star position is marked with a yellow asterisk. The FOV of each displayed image is $8.2'' \times 3.1''$. \label{f_imaging}    }
 \end{figure*}

\begin{figure*}
 \centering
 \includegraphics[width=17.5cm]{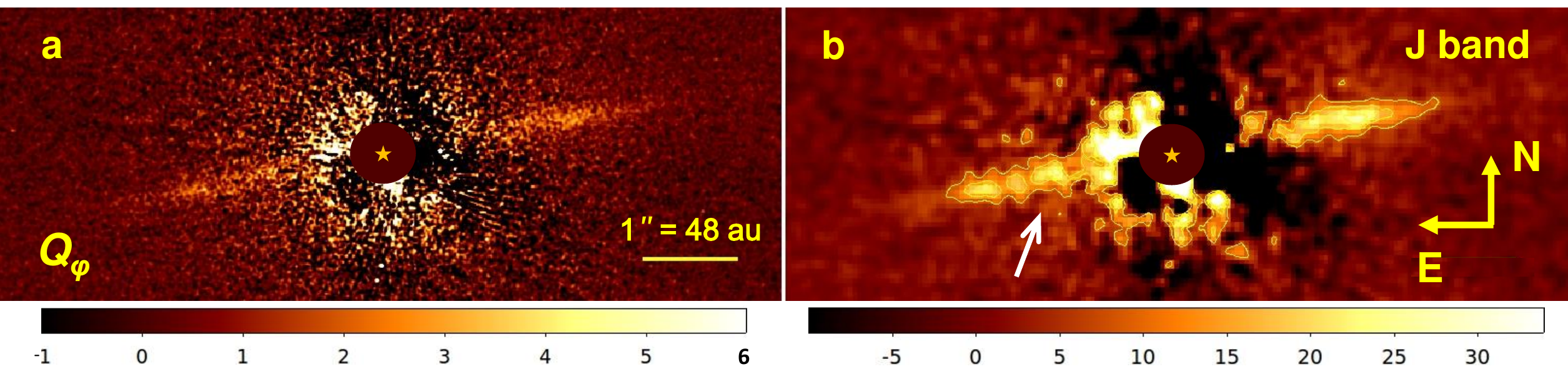}
    \caption{Polarized intensity images of the HD\,15115 debris disk in J band. Images show the Stokes $Q_\varphi$ parameter smoothed by a Gaussian kernel with $\sigma =$ 2 pixels ({\it a}) and with an additional $4\times 4$ binning ({\it b}). The white arrow points to the local surface brightness peak responsible for the impression of the inner belt (see also Fig.~\ref{f_disks}). The star position is marked with a yellow asterisk. The color scale shows the flux in counts per binned pixel. The surface brightness contours in image ({\it b}) are given for 5 and 15 counts per binned pixel. The FOV of each displayed image is $8.2'' \times 3.1''$.  \label{f_QphiUphi}}
 \end{figure*}
 
\section{Data reduction}\label{data}
\subsection{Classical imaging data} \label{s_Imaging}
To produce the calibration data, raw frames (flats, darks and sky background) are processed with the SPHERE data-reduction and handling (DRH) pipeline {\it esorex} \citep{Pavlov2008}. A custom routine is used to perform the basic cali- bration of the science frames: a bad pixels removal, the subtraction of the sky (instrumental) background and the flat-field correction. The calibrated science frames with the size of 2048$\times$1024 pixels are then split in two halves to create two temporal data cubes (the left and right channels of the instrument) containing 192 (H band) or 384 (J band) frames of 1024$\times$1024 pixels each. 

The position of the star is determined by fitting a 2D Gaussian function to the four satellite spots in the ``center frame''. The star coordinates are derived from the intersection point of two lines connecting the centers of two opposite spots. These coordinates are used for re-centering all frames in the data cube.

In the next step, reduced and centered data cubes are processed with several custom algorithms making use of ADI technique: cADI \citep[classical ADI;][]{Marois2008}, KLIP \citep[Karhunen-Lo\`{e}ve Image Projection;][]{Soummer2012} and LOCI \citep[Locally Optimized Combination of Images;][]{Lafreniere2007}. 

The final median-combined images obtained with cADI, after the stellar light had been removed, are shown in Figs.~\ref{f_imaging}a and \ref{f_imaging}d. 

Similar results were obtained with the KLIP algorithm (Figs. \ref{f_imaging}b and \ref{f_imaging}e). This method uses a principal component analysis to construct a reference image for the stellar light from the projection of the science image onto an orthogonal basis given by the Karhunen-Lo\`{e}ve (KL) transform of the same science images. 
We performed KLIP data reductions with 3, 5, 15, and 25 KL vectors (or eigenimages) forming the basis for a reference image. The entire images were used without a frame selection to avoid large disk-flux self-subtraction effects. Visual inspection of the final images showed that the larger number of vectors yields a very strong self-subtraction of the disk flux. 

For comparison, we also performed the LOCI data reduction (Figs. \ref{f_imaging}c and \ref{f_imaging}f). The LOCI algorithm provides a high detectability performance similar to the KLIP procedure. The reference image is constructed for each frame individually as a linear combination of other images in the data cube which have a minimal difference in the parallactic angle to the processed frame. The requested minimum difference in field rotation, in units of arc length, is set by parameter $N_\delta$. The optimization algorithm works in annular segments with the size specified by the parameter $N_A$ which gives the number of ``PSF cores'' that should fit in the segment area. The frames must cover a large rotation angle to preserve the flux of an extended structure like an edge-on debris disk. For the LOCI reduction we adopted parameters $N_\delta=1.25\times$FWHM and $N_A$ equal to 1000 and 300. 

To correct for the IRDIS True North (TN) offset, all frames are additionally rotated clockwise by $1.7^{\circ}$ \citep{Maire2016} during the ADI procedure.

Reduced frames of the stellar flux measurement are centered by fitting a 2D Gaussian function with a FWHM of $\sim$3 pixels to the stellar profile. The sky subtraction is not applied to these frames. Instead, we measure the instrumental background in the median frames of both IRDIS channels and subtract it.
The median frames are then divided by the mean transmissivity of the neutral density filter ND2.0 equal to $3.3 \cdot10^{-3}$ and $1.5 \cdot10^{-2}$ in the J and H bands, respectively (see SPHERE manual\footnote{\url{https://www.eso.org/sci/facilities/paranal/instruments/sphere/doc.html}}). The final frames that are used to measure the stellar count rate (see Sect.~\ref{s_Photometry}) are the mean frames of two flux measurements performed before and after each observation. \\\
 
\subsection{Polarimetric data} \label{s_polarim}
We used the DRH pipeline {\it esorex} to produce the calibration data. The raw polarimetric science frames are calibrated with the same custom procedure (subtraction of the sky or instrumental background, flat-field correction, bad pixel removal) as applied to the total intensity data.

The Stokes $Q$ and $U$ parameters were obtained with the double-difference method from calibrated data using a custom routine. 
We applied the same basic data-reduction steps as described in \citet{Engler2017} where the procedure is explained in detail.
For the frame centering we have used the same method as for the centering of the total intensity data set (see Sect. \ref{s_Imaging}). After that, each $Q$ and $U$ frame was visually inspected to check the correct position of the center of the right channel frame relative to the  left channel frame. 

The final Stokes $Q$ and $U$ parameters are calculated as follows:
\[Q = 0.5*(Q^+ - Q^-)\]
\[U = 0.5*(U^+ - U^-),\]
where $Q^+$ is measured along the axis pointing to the north, $U^+$ is measured along the axis pointing to the northeast, and $Q^-$ and $U^-$ are measured along the axes perpendicular to $Q^+$ and $U^+$, respectively. 

The $Q$ and $U$ images contain the linear polarization components of the total polarized flux received from the debris disk. To generate the image of the disk showing the total polarized flux, the instrumental coordinate system is transformed to the polar coordinate system where the linear polarization components are represented by the azimuthally oriented Stokes $Q_\varphi$ parameter and $U_\varphi$ parameter, the linear polarization component at a 45$^\circ$ angle to the $Q_\varphi$ parameter:
\[Q_\varphi = - Q\cos 2\varphi-U\sin 2\varphi\]
\[U_\varphi = Q\sin 2\varphi - U\cos 2\varphi\]
where $\varphi$ is the polar angle measured east of north in the coordinate system centered on the star. Here we adopt the sign convention for $Q_\varphi$ and $U_\varphi$ parameters which is consistent with $Q_{\rm r}$ and $U_{\rm r}$ parameters defined in \citet{Schmid2006}.

The instrumental polarization signal is corrected through a forced normalization of the total counts in frames of opposite polarization states. 

Finally, the $Q_\varphi$ and $U_\varphi$ images are rotated clockwise by $1.7^{\circ}$ to perform the astrometric TN correction (see Sect.~\ref{s_Imaging}).

Figure \ref{f_QphiUphi} shows the $Q_\varphi$ image with the detected polarized light. The debris disk appears as a thin almost symmetric stripe oriented in the east-west direction and extending beyond 2.5$''$ (120 au). There is no detection in the $U_\varphi$ image, as expected. Therefore the $Q_\varphi$ image shows the total intensity of the polarized light. 

\begin{figure*}[ht]
   \centering
   \includegraphics[width=17cm]{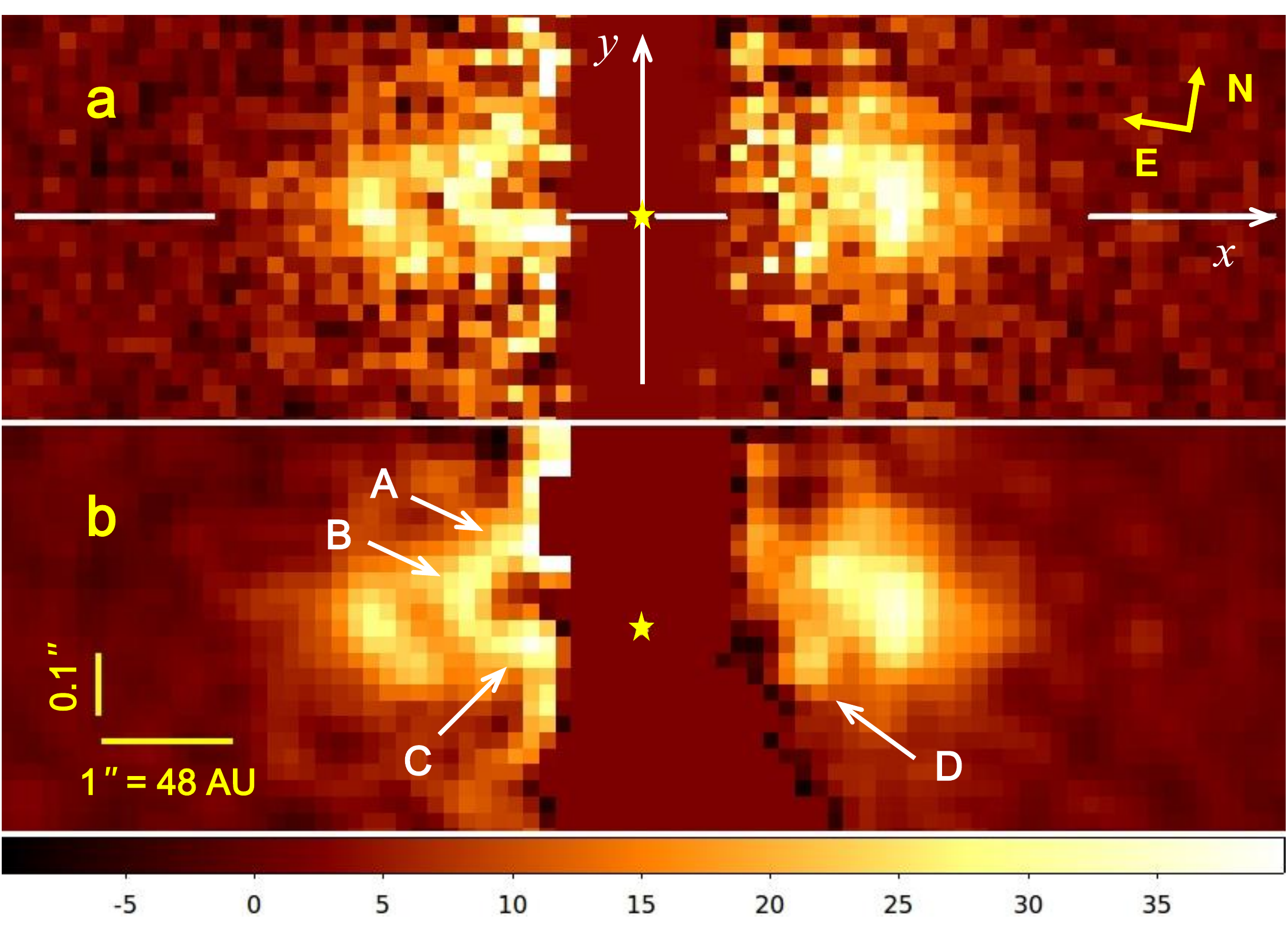} 
   \caption{$Q_\varphi$ images binned by 10 pixels along the $x$-axis and by 2 pixels along the $y$-axis and smoothed via a Gaussian kernel with $\sigma =$ 1 pixel ({\it a}) and with $\sigma =$ 2 pixels ({\it b}) to highlight the structure and position of both rings relative to each other. The central part of the images, dominated by strong residuals, is masked out. The position of the star is indicated by a yellow asterisk. The $x$-axis (interrupted white line going through the star position) shows the outer disk major axis at PA = 279.0$^\circ$. The FOV of the displayed image ({\it a}) is $9.8'' \times 0.735''$. The letters 'A', 'B', 'C' and 'D' are used to label the surface brightness peaks discussed in Sect.~\ref{Morphology}. Color bar shows flux in counts per binned pixel.  \label{f_disks}}
  \end{figure*}

\section{Data analysis} \label{Data Analysis}
\subsection{Disk morphology} \label{Morphology}   

\subsubsection*{Total intensity data}
All data reductions (cADI, KLIP, LOCI) reveal an almost edge-on disk with the bright side to the north of the disk major axis. The famous bow-like structure, previously reported in different studies \citep{Rodigas2012, Mazoyer2014, Schneider2014} and corresponding to the brightest side of the highly inclined ring, is clearly visible in both J and H. Additionally, our images unveil the disk ansae symmetrically located to the east and west of the star. In some images, parts of the fainter side of the disk are also visible near the ansae. The brighter disk side is detected with a signal-to-noise ratio (S/N) higher than 4 (see S/N map in Appendix, Fig.~\ref{f_imaging_snr}) up to a radial distance of $\sim$4$''$ eastwards and $6''$ on the west side (up to the end of frame).  

\subsubsection*{Polarized intensity data}
Figure \ref{f_disks} shows the binned $Q_\varphi$ image after binning by 10 and 2 pixels along the $x$- and $y$-axis, respectively. The smaller vertical binning is applied to preserve the spatial information provided by the high-resolution imaging. In this figure we define an $x-y$ coordinate system (Fig. \ref{f_disks}a) where the star is at the origin, $+x$ is the coordinate along the major axis westwards ($\theta_{\rm disk}$), $-x$ is eastwards ($\theta_{\rm disk}-180^\circ$), and the $y$- axis is perpendicular to this with the positive axis towards the north.

There are three surface brightness peaks on the east side within $r=1.5''$ in azimuthal direction: at A$(-0.80'',\, 0.29'')$, B$(-1.30'',\, 0.05'')$, and C$(-0.75'',\, -0.16'')$ indicated in Fig. \ref{f_disks}b. However, the points A and C are inside the AO correction area dominated by strong stellar residuals and cannot be considered as a detection. A straight line going through point B and the star intersects the small region with increased surface brightness on the west side shown by arrow D. This region is mirror-symmetrically located to the point B with respect to the star. 

\begin{figure*}
   \centering
   \includegraphics[width=17cm]{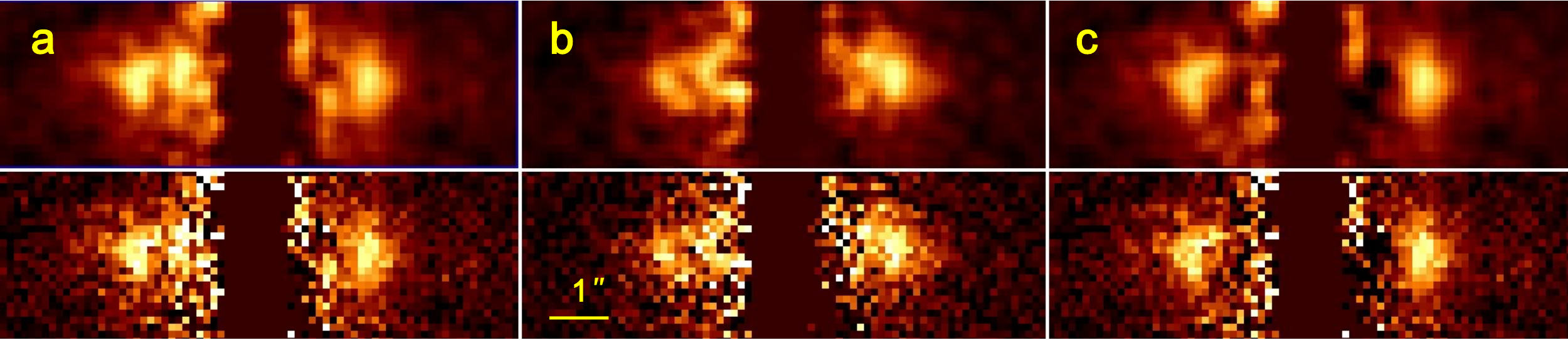} 
   \caption{Comparison of the double (\textit{column a}) and single (\textit{column c}) belt models with the $Q_\varphi$ image (\textit{column b}). All images are binned by 10 pixels along the $x$-axis and by 2 pixels along the $y$-axis and additionally smoothed via a Gaussian kernel with $\sigma =$ 2 pixels (\textit{top row}) and with $\sigma =$ 1 pixel (\textit{bottom row}).  \label{f_Models_intro}}
  \end{figure*}

Such a distribution of the polarized surface brightness conforms with the highly inclined double belt system in which the outer and inner belts have slightly different PA and inclination from our toy model of the system. Figure~\ref{f_Models_intro} shows the synthetic images of the polarized intensity in a debris disk with single and double belt in comparison to the data. To create synthetic images, the model\footnote{The detailed description of the model used can be found in Sect.~\ref{s_modeling}. } with radius of the disk $R_0= 2''$ and inclination of $86^\circ$ was used. The other parameters were arbitrarily chosen, for instance, the radius of the inner belt was set to $1.5''$. In order to have a similar level and pattern of noise in the simulated images and in the data, both models (single and double belt) were inserted into the $Q_\varphi$ image. The disk flux was previously removed from this $Q_\varphi$ image. This was done by replacement of the image area containing disk flux through the same area of the mean of two images obtained by rotation of the $Q_\varphi$ image by $15^\circ$ clockwise (first image) and by $15^\circ$ counterclockwise (second image). The model images were binned and smoothed in the same way as was done to obtain the $Q_\varphi$ image presented in Fig.~\ref{f_disks}. The visual comparison between model images (Fig.~\ref{f_Models_intro}a and \ref{f_Models_intro}c) and the $Q_\varphi$ image (Fig.~\ref{f_Models_intro}b) suggests that the HD\,15115 debris disk could be a double-ring system because the SB distribution observed in particular on the east side of the star can be explained by the presence of a second inner belt. We also note that the single-belt model (Fig.~\ref{f_Models_intro}c) lacks polarized flux inside 1.5$''$.  

Motivated by this comparison, we fitted an ellipse to the SB peaks A, B, C, and D and found that, if such an inner belt were to exist, it would possibly have a deprojected radius of $r\approx 1.25'' \pm 0.07''$ and a PA $\theta_{\rm inn\, ring} = 276^{\circ} \pm 3^{\circ}$, and thus a difference in PA up to 6$^{\circ}$ with respect to the outer belt.

The goal of the data analysis presented in the following sections is to find out whether there is some additional evidence for the double-belt structure suspected from the polarimetric observation. For this purpose, we explore the position of the disk spine (Sect.~\ref{s_Spine}), compare the models for the spatial distribution of dust in an optically thin disk with the total and polarized intensity data (Sect.~\ref{s_modeling}), and investigate the variation of the disk surface brightness along the disk axis calculated in apertures of different sizes (Sect.~\ref{s_Photometry}).

\subsection{Position angle and size of the disk} \label{PA_size}
We estimated the disk radius, the PA, and the inclination from the H band imaging data reduced with LOCI and KLIP algorithms where the disk front and back sides are visible. We derived the $R_{\rm disk}= 2.01'' \pm 0.05''$, $\theta_{\rm disk}=278.7^{\circ} \pm 0.2^{\circ}$ and the disk inclination $i = 86.5^{\circ} \pm 0.5^{\circ}$ by fitting an ellipse to the ring contours and the surface brightness peaks at the east and west disk edges. 

We also derived the PA and the radius of the disk by determining the centroids of bright ansae
at the radial separation $r\approx 2''$ to the east and west of the star in the polarimetric data in J band (Fig. \ref{PA}). This location corresponds to the radius of the symmetrical ring $R_{\rm disk}= 1.99''$ detected by \cite{Mazoyer2014} in the H band data which is evident in our data as well. For an edge-on disk or ring with a characteristic radius $R_{\rm disk}$, the maximum polarized flux is expected at the locations of the planetesimal belt or at scattering angles close to $90^{\circ}$. This is a natural outcome of the interplay between the angular distribution of the polarized flux, which is described by the polarimetric phase function, and the radial distribution of the grain number density, which is maximal at $R_{\rm disk}$ (see also Sect.~\ref{s_pol_model}). This provides a relatively simple method to directly measure the extent of the planetesimal belt and its position on the sky. Using this approach we obtain for the HD\,15115 outer disk the radius $R_{\rm disk} = 1.96'' \pm 0.05''$ ($94.39 \pm 2.41$ au) along the disk major axis and the PA $\theta_{\rm disk}=279.0^{\circ} \pm 0.2^{\circ}$. After taking into account the PA measured in the total intensity data, this yields a mean PA of the disk $\theta_{\rm disk}=278.9^{\circ} \pm 0.1^{\circ}$.

These values are in good agreement with the disk radius $R_{\rm disk} = 1.99''$, PA $\theta_{\rm disk}=278.8^{\circ} \pm 0.4^{\circ}$ and the disk inclination $i = 86.2^{\circ}$ found by \cite{Mazoyer2014} in the NICI data. Our measurement of the PA agrees also very well with results from previous studies of HD\,15115 at different wavelengths \citep{Kalas2007, Schneider2014}. 
We do not see any evidence in our data of an east/west misalignment as noted by \cite{Debes2008} who obtained PA $\theta_{\rm disk} = 278.9^{\circ} \pm 0.2^{\circ}$ for the west side and PA $\theta_{\rm disk} = 276.6^{\circ} \pm 1.3^{\circ}$ for the east side at 1.1 $\mu$m. It should be noted that the NICMOS data have much lower resolution compared to the IRDIS data and the PA measurement of \cite{Debes2008} could be affected by the strong stellar residuals around the coronagraph. \\

\begin{figure*}  
   \centering
   \includegraphics[width=15cm]{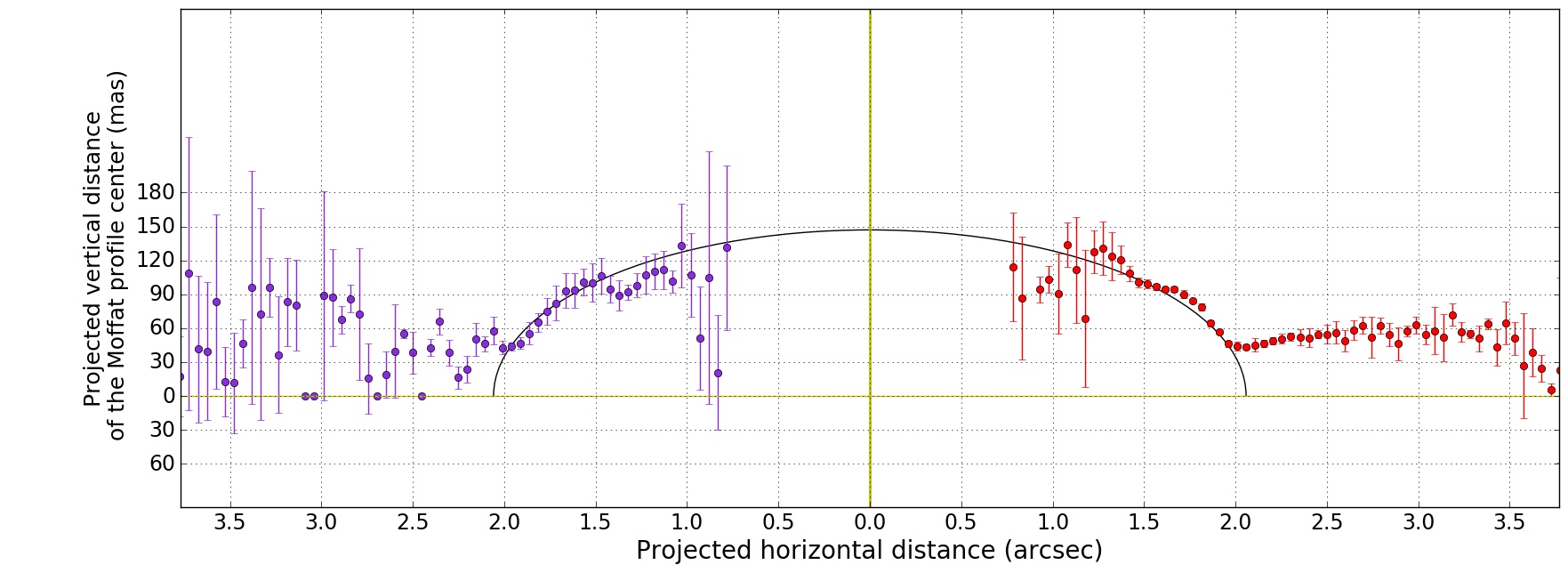}
   \caption{Spine vertical distance from the disk major axis measured as a vertical offset of the Moffat profile center. The plot traces the mean location of the spine in all available images (cADI, KLIP, LOCI reductions of the total intensity data) in H and J bands. The star position is at (0,~0). The black line shows the best-fit ellipse to the data points between $|x| \geq 1.2''$ and $|x|\leq 2.0''$. \label{f_Spine}}
\end{figure*}

\subsection{Spine of the disk} \label{s_Spine}   
The geometric parameters of the disk can be inferred from the spine which we define as the location of the maximum intensity of a Moffat function fitted to the disk cross-sections perpendicular to the midplane. The Moffat function is given by the following equation \citep{Trujillo2001}:
\[f_M(y)=a_M\left[ 1+ \left(\frac{y-y_0}{\alpha}\right)^2\right] ^{- \beta},\]
where $a_M$ is the peak flux located at a vertical distance $y_0$ from the disk major axis and $\alpha$ and $\beta$ are two constant parameters determined from the fit.

We measured the spine position of the brighter side of the disk for several data reductions (cADI, KLIP, LOCI) in H and J bands (with a 4$\times$4 binning) and derived the mean vertical offset $\left\langle y_0(x)\right\rangle $ as plotted in Fig. \ref{f_Spine}. The $y_0$ values, which are larger than $y_0 = 100$ mas or smaller than $y_0 = -25$ mas, are considered as poor fits due to a low S/N and are excluded from this analysis. The error bars show the standard deviation of the offset values obtained in all data sets without taking these outliers into account.

The spine has a well defined bow-like shape as already described in \cite{Mazoyer2014} (see their Fig. 5) but here with a much smaller dispersion. The smaller error bars are due to the improved angular resolution and contrast provided by SPHERE compared to NICI. We also note that the spine dispersion is relatively small for the radial separations between $1.2''$ and $3.7''$ on the west side and between $1.0''$ and $2.4''$ on the east side.

Between $r\approx1.2''$ and $r\approx2.2''$ the spine is roughly symmetric with respect to $ x_0$. The smallest $\left\langle y_0(x)\right\rangle $-offset is approximately $45\pm 5$ mas (2.2 au) around $|x|= 2.0''$ $\pm 0.1''$. For smaller radial separations the vertical offset is increasing and achieves a maximum of $\sim$140~mas at $x\approx \pm 1.0''$ in our data. For radial separations $x > 2.0''$ the spine deviates from the major axis by 45--60\,mas towards the north.
The same behavior is observed on the east side, although with a lower S/N. A similar spine curve which does not reach the disk major axis at the location of the planetesimal belt was measured for another edge-on debris disk around F star HIP 79977 \citep{Engler2017} and A star $\beta$ Pictoris \citep{Milli2014}. Such a course of the spine curve is expected (see Sect.~\ref{s_Spine_diagnosticsII}) and explained by the anisotropic scattering of the dust \citep{Mazoyer2014}. 
\paragraph*{\bf {Spine diagnostics I}} The turning point of the spine at $\sim$2$''$ clearly indicates the location of the edge of the parent belt of the planetesimals. The geometry of the disk bow traced by the accurately measured spine allows us to put constraints not only on the characteristic radius of the belt but also on the disk inclination. We derive these parameters by fitting an ellipse to the spine between $|x| \geq 1.2''$ and $|x|\leq 2.0''$. The center of the ellipse is set to the star position. The best-fit ellipse (Fig. \ref{f_Spine}) has a semi-major axis of $2.06'' \pm 0.06''$ and a semi-minor axis of $0.15'' \pm 0.01''$ yielding a disk inclination of $85.8^{\circ} \pm 0.7^{\circ}$. This inclination estimate is in good agreement with $i=86.2^{\circ}$ found by \cite{Mazoyer2014} with the same methodology.

\section{Disk modeling} \label{s_modeling} 
\subsection{One belt model} \label{s_one_belt}  
To put constraints on the basic geometric parameters of the disk, we generate a set of models from a parameter grid and compare the synthetic images for the scattered light with the KLIP disk images in J and H bands (Figs. \ref{f_imaging}b and \ref{f_imaging}e) by evaluating the reduced $\chi^2_\nu$ for the model fit (Eq.~\ref{eq_chi2}). 

The synthetic image of scattered light is created for each model assuming a rotationally symmetric distribution of the dust. In accordance with the theory of the ``birth'' ring, small dust grains are generated in a collisional cascade triggered by large planetesimals residing in a narrow ring at a distance $R_0$ from the star. Therefore, the radial distribution of the grain number density has a peak at the location of the parent belt and can be described by the power laws $r^{\alpha_{\rm in}}$ $(\alpha_{\rm in}>0)$ for $r<R_0$ and $r^{\alpha_{\rm out}}$ $(\alpha_{\rm out}<0)$ for $r>R_0$ to account for the decrease of the grain number density 
inside and outside of the ring \citep[e.g.][]{Augereau2001}. Adopting a Gaussian function for a vertical distribution, the grain number density distribution $n(r, h)$ in the disk can be written in cylindrical coordinates $(r, h)$ as:
\begin{equation} \label{R}
n(r, h) \sim {\left( \left(\frac{r}{R_0}\right)^{-2\alpha_{in}}+\\\ \left(\frac{r}{R_0}\right)^{-2\alpha_{out}} \right)}^{-1/2} \cdot \exp\left[- \,{\left(\frac{|h|}{H(r)}\right)}^2 \right],
\end{equation}
where $H(r)$ is the disk scale height which scales with a radius like $H(r)= H_0 \,{\left(\frac{r}{R_0}\right)}^{\gamma}$, $H_0$ is the scale height of the vertical profile of the disk at $R_0$, and $H_0 / R_0$ defines the aspect ratio of the disk. For simplicity, we set the disk flare index to $\gamma = 1$.

Further, we assume that the disk is optically thin, that is, that the single scattering dominates over multiple scattering of photons, and adopt the Henyey-Greenstein (HG) phase function for the angular dependence of the scattered radiation \citep{Henyey1941}:
\begin{equation} \label{HG}
f(\theta)=\frac{1-g^2}{4 \pi (1+g^2-2 g cos (\theta))^{3/2}},
\end{equation}
where $g$ is the average of the cosine of the scattering angle for the phase function. There is $g=0$ for the isotropic scattering, forward scattering grains have $0<g\leq 1$, while for $-1\leq g<0$ the scattering is peaked backwards. In this work, we do not consider $g< 0$ assuming only forward scattering for the dust grains. We do not introduce the size distribution of dust grains to calculate the asymmetry parameter $g$ and the cross section for scattering $\sigma_{\rm sca}$ averaged over all particle sizes \citep{Engler2017}. In our model, these parameters are constant through the disk, which is a reasonable approximation if we take into account that disk images preferentially probe grain sizes comparable to the wavelength of observation.

%%%%%%% Tables with modeling results
 \begin{table*}  
      \caption[]{Best-fit model parameters of the H band data. \label{t_resultsH}}
 % v4
	     \centering
         \begin{tabular}{lcccccc}
            \hline
            \hline
            \noalign{\smallskip}
       Optimized &\multicolumn{2}{c}{East side ($\nu =425$)} & \multicolumn{2}{c}{West side ($\nu =635$)} & \multicolumn{2}{c}{Both sides ($\nu =1065$)}\\     
 	    parameter& Best fit& Models with &Best fit&Models with &Best fit$^1$& Models with\\
 	     	     & $\chi^2_{\nu,\,{\rm min}}=1.40$ &$\chi^2_{\nu}<1.47$ & $\chi^2_{\nu,\,{\rm min}}=1.53$ &$\chi^2_{\nu}<1.58$  & $\chi^2_{\nu,\,{\rm min}}=3.63$ &$\chi^2_{\nu}<3.67$\\
            \hline
            \noalign{\smallskip}
       HG parameter $g$                  & $0.5$ & 0.5, 0.6 & 0.3 & 0.3 & 0.3 & 0.3-0.4\\[5pt]
	   Inner radial index $\alpha_{\rm in}$    & 9   & 6-9       & 2   &2-4 & 2   &2-3\\[5pt]
	   Outer radial index $\alpha_{\rm out}$ & -6 & -6      &-4   & -4 & -4 & -4\\[5pt]
       Aspect ratio $H_0 / R_0$        & 0.05 & 0.03--0.05 & 0.01 & 0.005--0.03 & 0.01 & 0.005--0.03\\[5pt]
	   \noalign{\smallskip}
           \hline
            \hline
            \noalign{\smallskip}
         \end{tabular}
\begin{flushleft}
{\bf Notes.} $^{(1)}$ The best-fit model with $\chi^2_{\nu}=3.38$ is the two belts model with an inner belt inclined at  $i=80^{\circ}$ (for parameters see Sect. \ref{s_two_belts}). In Col. 6, the parameters of the second best fitting model with only one belt are given. \\
\end{flushleft}
\end{table*}

 \begin{table*}  
      \caption[]{Best-fit model parameters of the J band data. \label{t_resultsJ}}
	     \centering
         \begin{tabular}{lcccccc}
            \hline
            \hline
            \noalign{\smallskip}
       Optimized &\multicolumn{2}{c}{East side ($\nu =425$)} & \multicolumn{2}{c}{West side ($\nu =635$)} & \multicolumn{2}{c}{Both sides ($\nu =1065$)}\\     
 	    parameter& Best fit& Models with &Best fit&Models with &Best fit & Models with\\
 	     	     & $\chi^2_{\nu,\,{\rm min}}=1.67$ &$\chi^2_{\nu}<1.74$ & $\chi^2_{\nu,\,{\rm min}}=2.22$ &$\chi^2_{\nu}<2.27$  & $\chi^2_{\nu,\,{\rm min}} =4.24$ &$\chi^2_{\nu}<4.28$\\
            \hline
            \noalign{\smallskip}
       HG parameter $g$                  & 0.4 & 0.4, 0.5 & 0.3 & 0.3 & 0.3 & 0.3\\[5pt]
	   Inner radial index $\alpha_{\rm in}$  & 3 & 2-6       & 3   &2-5 & 2   &2-3\\[5pt]
	   Outer radial index $\alpha_{\rm out}$ & -5 & -5, -6         &-3   & -3 &-3   & -3\\[5pt]
       Aspect ratio $H_0 / R_0$        & 0.03 & 0.03--0.05 & 0.03 & 0.01--0.05 & 0.03 & 0.075--0.03\\[5pt]
	   \noalign{\smallskip}
           \hline
            \hline
            \noalign{\smallskip}
         \end{tabular}
\end{table*}

The parameter space investigated in this work is based on the results of a previous modeling of the HD\,15115 disk and includes the following parameter sets (a grid of 2160 models in total):
\begin{itemize}
\item power-law index $\alpha_{\rm in}$: 2, 3, 4, 5, 6, 7, 8, 9, 10
\item power-law index $\alpha_{\rm out}$: -2, -3, -4, -5, -6, -7, -8
\item asymmetry parameter $g$:  0.2, 0.3, 0.4, 0.5, 0.6 
\item disk aspect ratio $H_0 / R_0$: 0.005, 0.0075, 0.01, 0.02, 0.03, 0.05.
\end{itemize}

The radius and inclination of the disk in all models are set to 2$''$ (96 au) and 86$^{\circ}$, respectively.

Each model is first rotated and inserted into an empty data cube to mimic the actual position of the disk in the data. Then each frame of this cube is convolved with the instrumental PSF measured from the data and the cube is processed through the KLIP algorithm. Thereby, we use the forward-modeling approach to account for the effects of the disk self-subtraction which always arise when the ADI technique is used \citep{Soummer2012}. In forward modeling, the synthetic images of models are projected on the same vector basis as determined during the processing of the original data cube. Figure \ref{f_1ring_model} shows an example of a synthetic model image (Fig.~\ref{f_1ring_model}a) convolved with the PSF (Fig. \ref{f_1ring_model}b) and then processed through the KLIP algorithm (Fig.~\ref{f_1ring_model}c) using an identical vector basis which was applied to obtain the KLIP images presented in Figs. \ref{f_imaging}b and \ref{f_imaging}e. The self-subtraction of the disk flux caused by the KLIP post-processing is clearly visible in the form of dark stripes on both sides of the disk in the data and the models. 

To estimate the goodness of the model fit to the KLIP images, we define the reduced $\chi_\nu^2$ metric as follows:
\begin{equation} \label{eq_chi2}
\chi^2_{\nu} = \frac{1}{\nu}\sum_{i=1}^{N_{\rm data}} \frac{\left[ F_{i,\, {\rm data}}- A\cdot F_{i,\, {\rm model}}(\vec{p})\right]^2 }{ \sigma_{i,\, {\rm data}}^2}
\end{equation}
where $N_{\rm data}$ is the number of binned pixels used to perform the $\chi^2_{\nu}$ minimization, $F_{i,\, {\rm data}}$ is the flux measured in pixel $i$ with an uncertainty $\sigma_{i,\, {\rm data}}$, $F_{i,\, {\rm model}}(\vec{p})$ is the modeled flux of the $i$ pixel and $A$ is a scaling factor. The degree of freedom of the fit is denoted by $\nu = N_{\rm data}-N_{\rm par}$, where $N_{\rm par}$ represents the number of variable parameters $\vec{p}=(p_{1}, p_{2}, ..., p_{N_{\rm par}})$ which is equal to five in this case. 

A $\chi^2_{\nu}$ metric is evaluated for each KLIP-processed model of the grid with respect to the data presented in Figs. \ref{f_imaging}b and \ref{f_imaging}e. We applied a $4\times 4$-pixel binning of the data and the models. 
Because of the pronounced brightness asymmetry between both disk sides, the west and east extensions were fitted separately in the first step. 
We performed the minimization in rectangular areas that are different for the west side ($1'' < x < 4''$ and $-0.2''< y < 0.2''$) and for the east side ($-3'' < x < -1''$ and $-0.2''< y < 0.2''$) for reasons related to the S/N. 
We also considered both disk sides simultaneously using the same areas. The error in intensity is calculated as the standard deviation of the flux distribution in binned pixels in concentric annuli excluding image areas which contain disk flux.   

We consider the model providing the minimum $\chi^2_{\nu, \, {\rm min}}$ as a best-fit model. Additionally, we determine the parameter sets of all well-fitting models bounded by a threshold of $\chi^2_{\nu}\leqslant \chi^2_{\nu, \, {\rm min}} + \Delta \, \chi^2_{\nu}$ \citep{Thalmann2014} to take the systematics into account. The  threshold is derived with $\Delta \, \chi^2_{\nu} = \sqrt{2/ \nu}$ corresponding to a 1$\sigma$ deviation of the $\chi^2$ distribution with $\nu$ degrees of freedom. Tables \ref{t_resultsH} and \ref{t_resultsJ} provide the best-fit model parameters (Cols. 2, 4, 6) as well as the range of parameters of well-fitting models satisfying the $\chi^2_{\nu}$ threshold noted in the column headers (Cols. 3, 5, 7). The parameter values of the best-fitting models can differ significantly between the east and west sides. 
The flux distribution on the west side can be well described with the HG asymmetry parameter $g=0.3 $ and the relatively shallow radial power-law indices $\alpha_{\rm in} = 3$ and $\alpha_{\rm out} = -4$ (H band data) or $\alpha_{\rm out} = -3$ (J band data), whereas the east side is best represented with $g =0.5$ and a much steeper index $\alpha_{\rm in} = 8$ (H band data) and $\alpha_{\rm out} = -6$. In the J band data, there is no large difference for the inner radial index $\alpha_{\rm in}$ between the east and the west side. The aspect ratio of 0.03 between disk radius and width seems to give a good fit to both disk sides, in particular when modeling J band data. 

As a final remark on the modeling results, we note that the $\chi^2$-minimization method is based on the assumptions that the flux measurement errors have a normal Gaussian distribution and the pixel values are independent. These assumptions might not be warranted in the considered data set and some other well-fitting parameter combinations might exist outside the quoted ranges.

\subsection{Two-belts model} \label{s_two_belts}  
The analysis of polarimetric data (Sect.\ref{Morphology}) suggests that the observed morphology of the HD\,15115 disk could be explained by the presence of an inner belt of planetesimals with a radius of $R< 2''$. If such an inner component exists it is not showing as an obvious feature in the total intensity data in particular because of the high inclination  of the HD15115 disk. The phase function in total intensity  being different than in polarimetry, an inner component will dominate the disk image at very small phase angles precisely where the ADI post-processing strongly attenuates the disk signal. To further test the disk geometry we consider both one- and two-belts model and compare their KLIP post-processed signatures.

\begin{figure}  
   \centering
   \includegraphics[width=8.5cm]{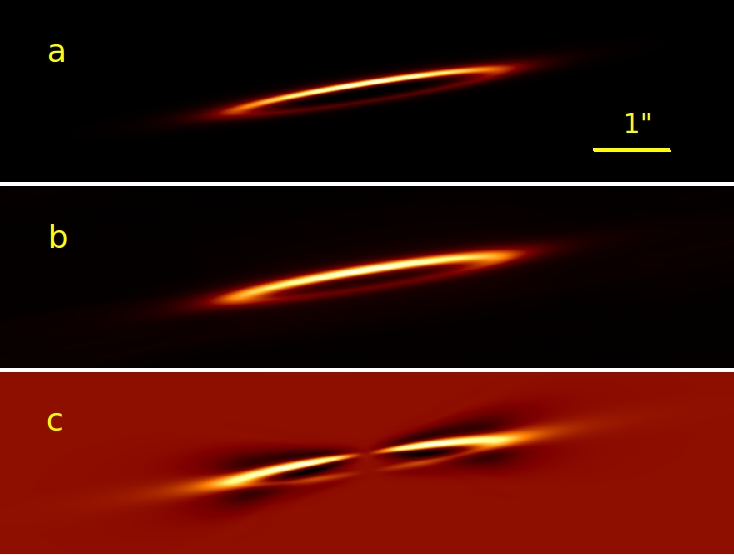}
   \caption{Example of synthetic images of one-belt model. {\it a}) Model before convolution with the instrumental PSF. {\it b}) Model convolved with the instrumental PSF. {\it c}) Model after processing with the KLIP algorithm. The flux is shown in arbitrary units.
 \label{f_1ring_model}}
\end{figure}

For this purpose, we use the one-belt model which gives a reasonable fit to the H band data ($\chi_\nu^2=1.81$ for the west side and $\chi_\nu^2=1.40$ for the east side). The model is generated with the following parameters: $R_0=2''$, $i=86^{\circ}$, $g=0.3$, $\alpha_{\rm in}=5$, $\alpha_{\rm out}= -4$, $H_0/R_0=0.01$ and position angle $\theta_{\rm out\, belt} = 279^{\circ}$. Figure~\ref{f_1ring_model} shows the appearance of the modeled belt (Fig.~\ref{f_1ring_model}a), the effect of the blurring instrumental PSF (Fig.~\ref{f_1ring_model}b) and the KLIP processing (Fig.~\ref{f_1ring_model}c) on it. 

For ease of comparison, when modeling the two-belts configuration, we use the same parameters for the outer ring as  described in the previous paragraph. The radius of the inner belt is set to $R_ {\rm inn\, belt} = 1.3''$ in accordance with the location of the surface brightness (SB) peaks (see Sect.~\ref{s_SBprofile}). For the remaining parameters of the inner belt, we adopt the mean values of the investigated parameter space: the asymmetry parameter $g=0.4$, radial power-law indices $\alpha_{\rm in}=5$ and $\alpha_{\rm out}= -5$, and the aspect ratio of 0.01. The intensity ratio between two belts is another free parameter in the two-belts model. 
For simplicity, we set this parameter to 1, assuming that the grain number density in the outer belt is higher by a factor of  $R^2_{\rm out\, belt} / R^2_{\rm inn\, belt}$ to take into account the light dilution with the distance from the star. For the analysis presented in this Section, we consider two models, each consisting of two belts: a non-coplanar model where the inner belt has a different inclination and PA from the outer belt, and a coplanar model where the inner belt has the same inclination and PA as the outer belt. 
\begin{figure}  
   \centering
   \includegraphics[width=8.5cm]{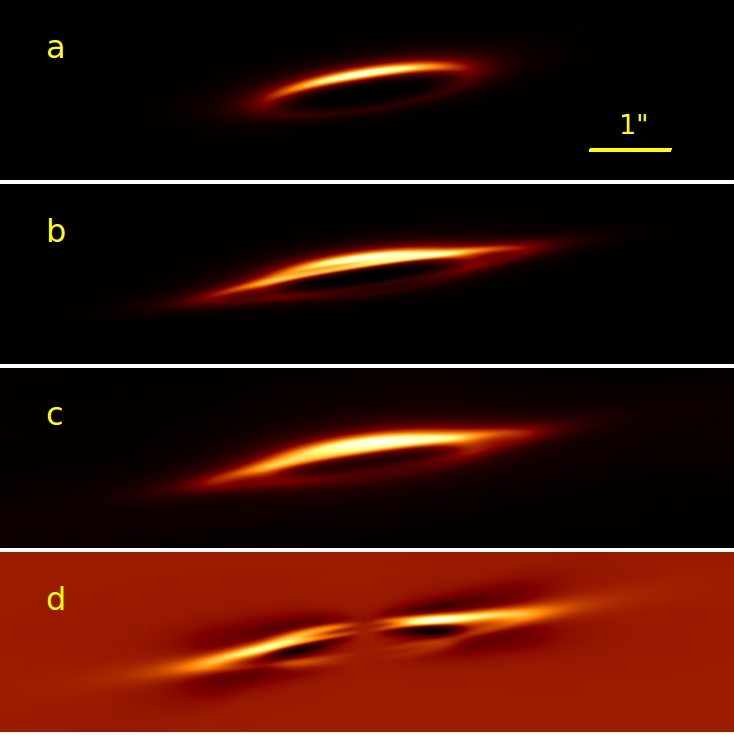}
   \caption{Synthetic images of the two-belts model with an inner belt (shown separately in image ({\it a})) which is non-coplanar with the outer belt. The outer belt has the same geometry and asymmetry parameter $g$ as the one-belt model demonstrated in Fig.~\ref{f_1ring_model}. Images ({\it b}) and ({\it c}) show the two-belts model before and after convolution with the instrumental PSF, respectively. Image ({\it d}) shows the KLIP-processed model. The flux is shown in arbitrary units.    \label{f_2_ring_model}}
\end{figure}
\begin{figure}  
   \centering
   \includegraphics[width=8.5cm]{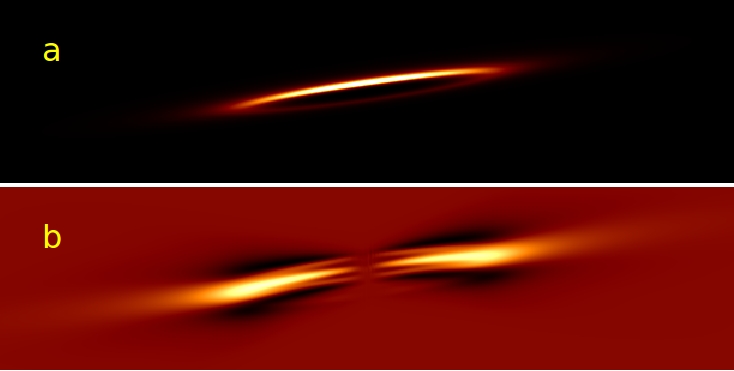}
   \caption{Synthetic images of the two-belts model with an inner belt coplanar with the outer belt. The model parameters are the same as in the model shown in Fig.~\ref{f_2_ring_model} except for the inclination and PA of the inner belt which are the same as for the outer belt. {\it a})~Model before convolution with the PSF. {\it b})~Model after processing with the KLIP algorithm. The flux is shown in arbitrary units. 
               \label{f_second_model}}
\end{figure}

In a non-coplanar model, the outer belt (PA $\theta_{\rm out\, belt}= 279^{\circ}$) has an inclination of $i=86^{\circ}$; the inner belt has a PA of $\theta_{\rm inn\, belt} = 276^{\circ}$ as measured from the polarimetric data (see Sect.~\ref{Morphology}) and an inclination of $i=80^{\circ}$. 
Such inclination was estimated by \cite{Schneider2014} for the inner ``half-ring'' component of the HD 15115 disk (see note to their Table~5). The inner ring is shown in the top panel of Fig.~\ref{f_2_ring_model}. The other panels of this figure present the non-coplanar model before convolution with the instrumental PSF (Fig.~\ref{f_2_ring_model}b) and after convolution (Fig.~\ref{f_2_ring_model}c), and the image obtained with the KLIP algorithm (Fig.~\ref{f_2_ring_model}d). 

The effect of a misalignment of two belts is clearly visible in Figs.~\ref{f_2_ring_model}b and \ref{f_2_ring_model}c. Because the relative PA and inclination of belts are non-zero, we expect the maximum SB of the disk to be shifted to the west side (to the right in this figure). The PSF convolution tends to blur the separation between the two belts (Fig.~\ref{f_2_ring_model}c), which however, is reinforced  afterwards (but also attenuated in intensity) owing to the high pass filtering produced by the KLIP processing (Fig.~\ref{f_2_ring_model}d).   

Figure~\ref{f_second_model} shows the coplanar two-belts model before convolution with the PSF (Fig.~\ref{f_second_model}a) and after KLIP processing (Fig.~\ref{f_second_model}b). The disk morphology in Fig.~\ref{f_second_model}a resembles that of a single ring (cf. Fig.~\ref{f_1ring_model}a) as expected. The two-belts model is slightly brighter in the central part and appears to be thicker in the KLIP-processed image (Fig.~\ref{f_second_model}b) compared to the one-belt model (cf. Fig.~\ref{f_1ring_model}c). 

The comparison between KLIP images of all models (Figs.~\ref{f_1ring_model}c, \ref{f_2_ring_model}d and \ref{f_second_model}b) illustrates the challenge of detection of multiple components in an edge-on disk. To see the split of the belts in a KLIP-reduced image, we need to resolve the system down to very small angular separations from the star. 

\begin{figure*}  
   \centering
\includegraphics[width=17cm]{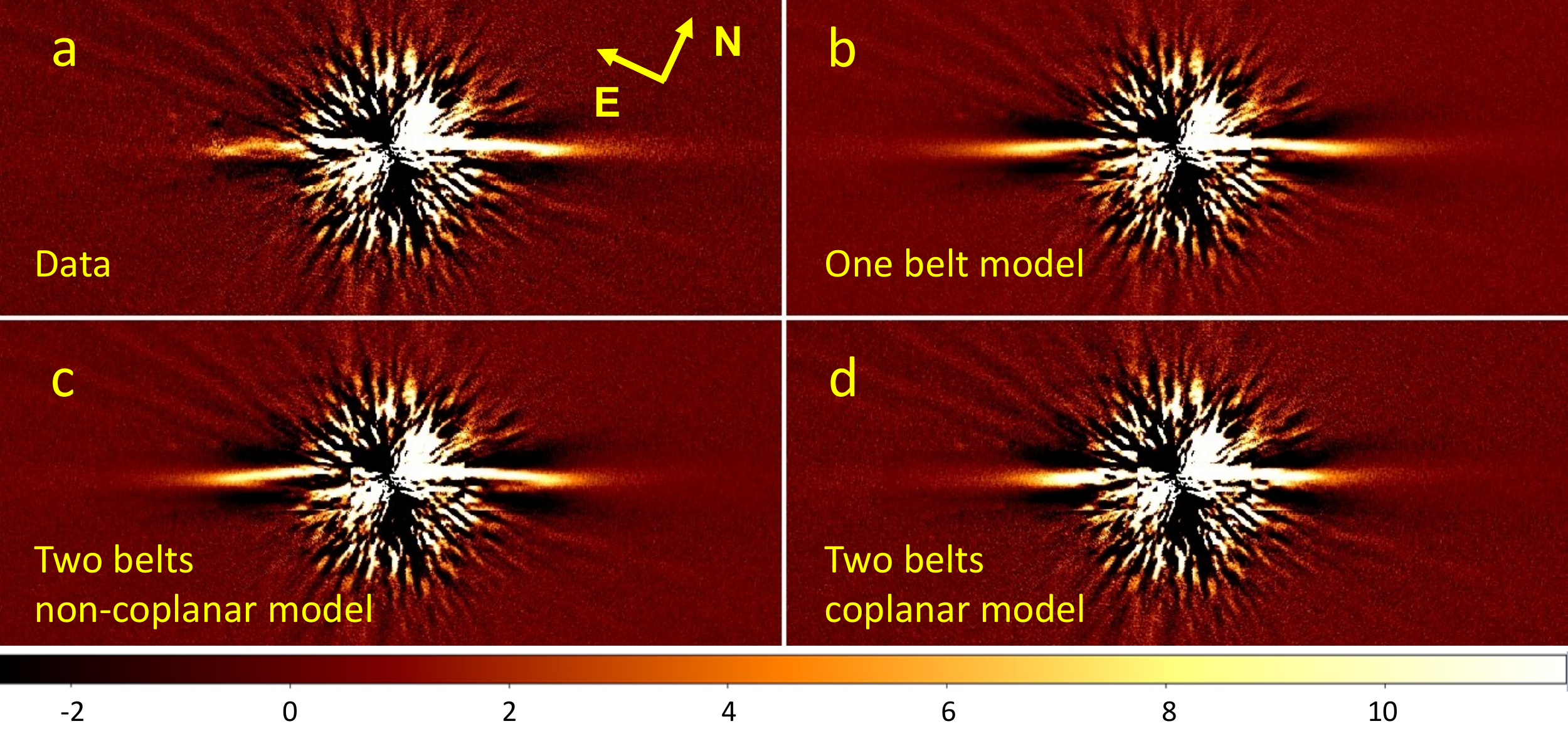}
       \caption{Comparison of the HD 15115 disk (\textit{panel a}) with the models inserted into the image (H band, KLIP reduction) where the disk was removed: \textit{b}) model with one belt; {\it c}) two-belts model with an inner belt which has a different PA and inclination with respect to the outer belt; {\it d}) two-belts model with an inner ring coplanar with the outer ring. In the processed images of the one-belt model (\textit{b}) and non-coplanar two-belts model (\textit{c}) the back side of the disk is visible, whereas it is not recognisable in the image of the  coplanar two-belts model (\textit{d}). Color bar shows flux in arbitrary units. \label{f_imaging_3models}}
\end{figure*}

In order to test how well the models with two belts fit the H band data, we calculated the reduced $\chi^2_{\nu}$ for the models presented in Figs.~\ref{f_2_ring_model}d and \ref{f_second_model}b setting the intensity ratio between the inner and outer belt equal to 0.5, 1, 2, 3 and~4. A simultaneous fit of both disk sides yields the $\chi^2_{\nu}$ in the range 3.38 and 3.67 for the non-coplanar model and in the range 3.68 and 3.89 for the coplanar model ($\nu=1060$, $N_{\rm par} = 10$). 
The minimum $\chi^2_{\nu}$ is obtained for the non-coplanar model with an intensity ratio of 3. This is the smallest value of all tested models (including single belt models) indicating a higher conformity with the data. The number of free parameters in this model is however larger than in the one-belt models, and it is possible that this is a result of data over-fitting. Therefore, we compare the best one-belt model (see Col.~6 in Table~\ref{t_resultsH} for the model parameters) with the non-coplanar two-belts model using the Bayesian information criterion ($BIC$) to check whether the additional parameters are warranted. This criterion introduces a penalty term for the additional parameters which increase the likelihood function $\mathcal{L}$ of the model $\mathcal{M}$. We define the $BIC$ as follows \citep{Schwarz1978}:
\begin{equation} \label{eq_BIC}
BIC(\mathcal{M}) = - 2 \ln(\mathcal{L}(\hat{\vec p})) +  N_{\rm par} \, \ln(N_{\rm data}),
\end{equation}
where $\hat{\vec p}$ is a set of the best-fitting parameters for the model $\mathcal{M}$ providing the maximized value of the likelihood function $\mathcal{L}$. Under the assumption that the model errors are normally distributed, this function can be estimated as:
\begin{equation} \label{eq_L}
\ln(\mathcal{L}) = - \frac{1}{2}\chi^2 + const
\end{equation}

The model with the lowest $BIC$ is preferred. We estimate the $BIC = 3901+ const$ for the model with one belt and $BIC = 3653+ const$ for the non-coplanar two-belts model. The latter $BIC$ is smaller, leading us to conclude that the double belt model provides a better fit to the data; additional parameters are justified. The change in $BIC$ of 248 is greater than 10, which is in support of the model with two belts \citep{Raftery1995}.
 
For all other variants of the fitting procedure (each side separately or using the J band data), the considered models with two belts are among the 20 best of all the models we tested. 
 
The visual comparison between the best-fitting one-belt model, the non-coplanar and coplanar two-belts models and the data is shown in Fig.~\ref{f_imaging_3models}. The models are inserted into the KLIP-processed H band image (Fig.~\ref{f_imaging_3models}a) where the disk flux was removed in the same way as was done with the $Q_\varphi$ image (see Sect.~\ref{Morphology}). Because all models are rotationally symmetric, their intensity is adjusted to match the intensity of the west disk side for an easier comparison with the asymmetric HD 15115 disk. The one-belt model (Fig.~\ref{f_imaging_3models}b) and the non-coplanar two-belts model (Fig.~\ref{f_imaging_3models}c) are comparable with the data (Fig.~\ref{f_imaging_3models}a). There is no significant visual difference between these two models, they both reveal the back side of the disk, for instance. On the contrary, the two-belts coplanar model (Fig.~\ref{f_imaging_3models}d) does not show the disk opening (back side is invisible). 

We conclude that the two-belts model with the inner ring parameters $R_ {\rm inn\, belt} = 1.3''$,  $i=80^{\circ}$ and PA $\theta_{\rm inn\, belt}= 276^{\circ}$ provides a better fit to the H band data than the coplanar model or single-belt model according to the $\chi^2_{\nu}$ method and the $BIC$. However, we have to keep in mind that this conclusion is based on the $\chi^2$ calculation and is therefore subject to the limitations mentioned in Sect.~\ref{s_modeling}. This means that none of the considered models can be absolutely ruled out. 
Higher spatial resolution and higher S/N of the data, especially at small angular separations, are necessary to differentiate between these models.

\subsection{Polarized light model} \label{s_pol_model}  
Since we have polarimetric data in the J band, we would like to test whether the two-belts model conforms with the polarimetric data set.

\begin{figure*}  
   \centering
\includegraphics[width=17.6cm]{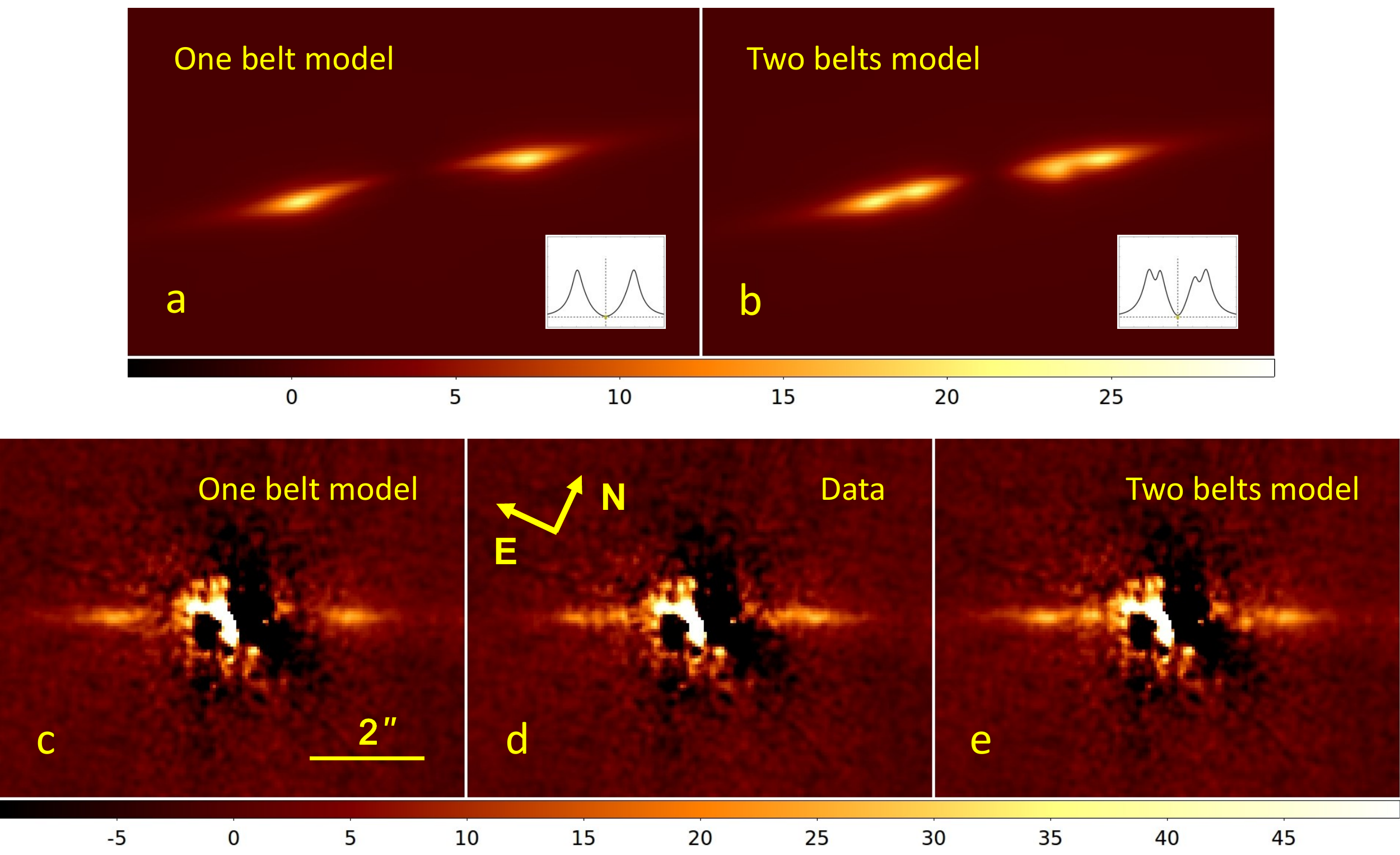}
               \caption{Model images of debris disk in polarized light ({\it top row}) and comparison of the $4 \times 4$ binned $Q_\varphi$ image ({\it d}) with one-belt model ({\it c}) and two-belts model ({\it e}) in the bottom row. The insets in the top row show the radial SB profiles of the one-belt model ({\it a}) and the two-belts model ({\it b}). All models are convolved with the instrumental PSF. Color bar shows flux in counts per binned pixel. \label{f_pol_model}}
\end{figure*}

Polarimetry is particularly useful to study the structure of an edge-on debris disk. Due to the different angular dependence of the scattered phase function and polarized phase function on scattering angle, the distribution of the disk SB and therefore the appearance of the disk are different in the images of scattered and scattered polarized light. For a nearly edge-on disk, the polarized disk flux has a maximum at scattering angles close to $\theta = 90^{\circ}$ \citep[e.g.][]{Engler2017} while the scattered flux peaks at $\theta = 0^{\circ}$ due to diffraction of light. The polarimetric measurement traces the extent of the planetesimal belt because we measure a peak polarized surface brightness of the disk at the location of the belt ansae. Therefore, in the case of the edge-on disk with multiple planetesimal belts, we should measure several SB peaks along the disk axis. These peaks indicate the presence of the inner belts and their extent.

In order to demonstrate this behavior, we simulated images of polarized light from an edge-on disk consisting of either one or two belts (Figs.~\ref{f_pol_model}a and \ref{f_pol_model}b). For these synthetic images we used the model described in \citet{Engler2017}. In this model, we assume that the degree of linear polarization of the scattered light $p_{\rm sca}$ has an angular dependence with the scattering angle $\theta$ such as that for Rayleigh scattering:
\[p_{\rm sca}(\theta) = p_m\frac{1- \cos^2\theta}{1+\cos^2 \theta},\]
where $p_m$ is the maximum fractional polarization achieved at a scattering angle of $\theta=90^\circ$.

The parameters of the one-belt model correspond to the best-fit model of the J band (see Col. 6 in Table \ref{t_resultsJ}). The two-belts model is the non-coplanar model shown in Fig. \ref{f_2_ring_model} (see Sect.~\ref{s_two_belts} for parameters). Both models were convolved with the PSF.

Comparison of Fig.~\ref{f_pol_model}a to Fig.~\ref{f_pol_model}b clearly demonstrates that polarimetric measurement helps to identify a multiple belt system. The polarized intensity image of such a system reveals several SB peaks symmetrically located with respect to the star on both disk sides. In this case, the radial SB profile measured along the disk major axis reflects the location and the number of the planetesimal belts (see insets in Figs.~\ref{f_pol_model}a and \ref{f_pol_model}b). 

Compared to the disk images in scattered light (see Figs.~\ref{f_2_ring_model}c or \ref{f_second_model}a), the polarimetric data offer the better opportunity to discover the inner components of the edge-on debris disk. In the total intensity images, the inner components are invisible. They can show up after the data post-processing, for example with the KLIP algorithm, but at very small radial separations from the star; they therefore remain inaccessible (Figs.~\ref{f_2_ring_model}d and \ref{f_second_model}b).

The polarimetric data of HD 15115 in the J band (Fig.~\ref{f_pol_model}d) could be interpreted as a two-belt system. However, for a meaningful conclusion concerning the disk structure based on the modeling results, a higher S/N of the polarimetric data is needed. Therefore we do not perform a fitting of the disk models to the $Q_\varphi$ image to explore the whole parameter space but compare qualitatively (visually) and quantitatively ($\chi^2_\nu$-value) only the one-belt model and the non-coplanar two-belts model discussed in Sect.~\ref{s_two_belts} with the polarimetric data (Fig.~\ref{f_pol_model}d). 

To do this, we inserted both models into the $Q_\varphi$ image which contains no disk flux and was already used for the illustration of the toy model (Sect.~\ref{Morphology}). The visual comparison of the resulting images reveals that the one-belt model (Fig.~\ref{f_pol_model}c) lacks the flux in the inner part of the disk in comparison to the data and to the two-belts model (Fig.~\ref{f_pol_model}e) and therefore the latter matches the $Q_\varphi$ image better.

Using the same areas as for the total intensity data and the same binning as well, we also found a marginal preference for the two-belts model with $\chi^2_\nu=3.49$ contrary to the one-belt model with $\chi^2_\nu=3.78$.
   
\subsection{Spine diagnostics II} \label{s_Spine_diagnosticsII}  

As discussed in Sect. \ref{s_Spine}, the spine measurement can be used as a diagnostic tool for the estimation of the disk radius and inclination. However, one should take into account the degeneracy between these parameters and other geometrical parameters describing the radial distribution of the dust such as $\alpha_{\rm in},\alpha_{\rm out}$, the HG asymmetry parameter $g$, and aspect ratio $H_0 /R_0$ \citep{Engler2017}. The course of the spine curve is defined by a combination of all these parameters. Therefore, in the following we investigate how the shape of the spine curve is affected if one of the parameters $\alpha_{\rm in}$, $\alpha_{\rm out}$, $g$, $H_0 / R_0$ or inclination varies. For this analysis, we use the KLIP processed grid models with the fixed radius of the belt ($R_0 = 2''$) as described in Sect.~\ref{s_modeling}.

\begin{figure*}  
   \centering
   \includegraphics[width=8cm]{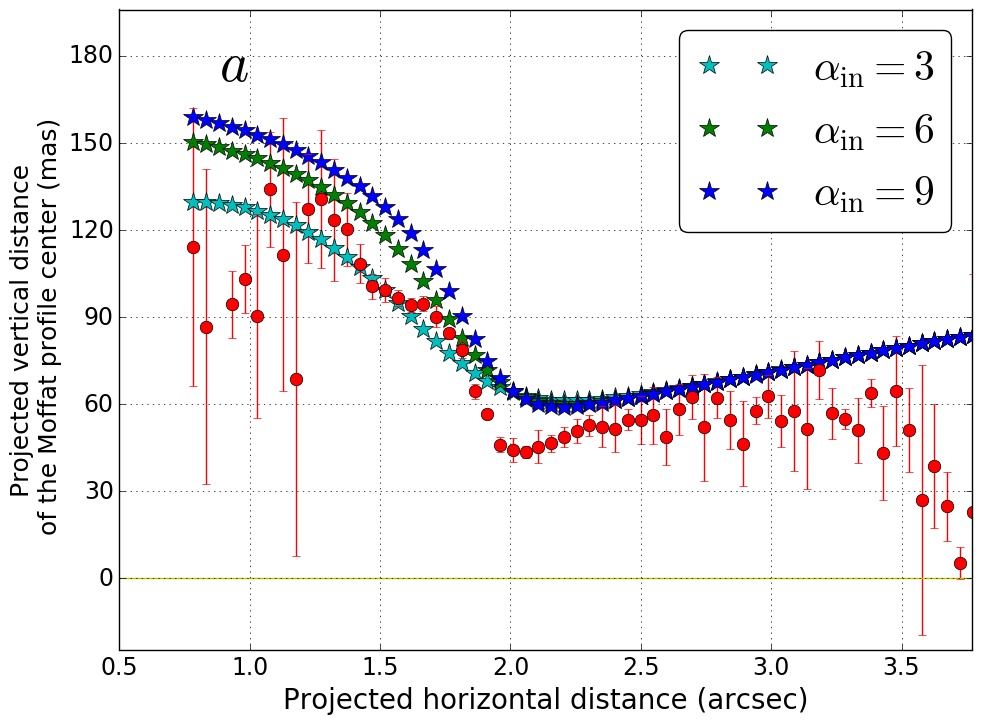}
   \includegraphics[width=8cm]{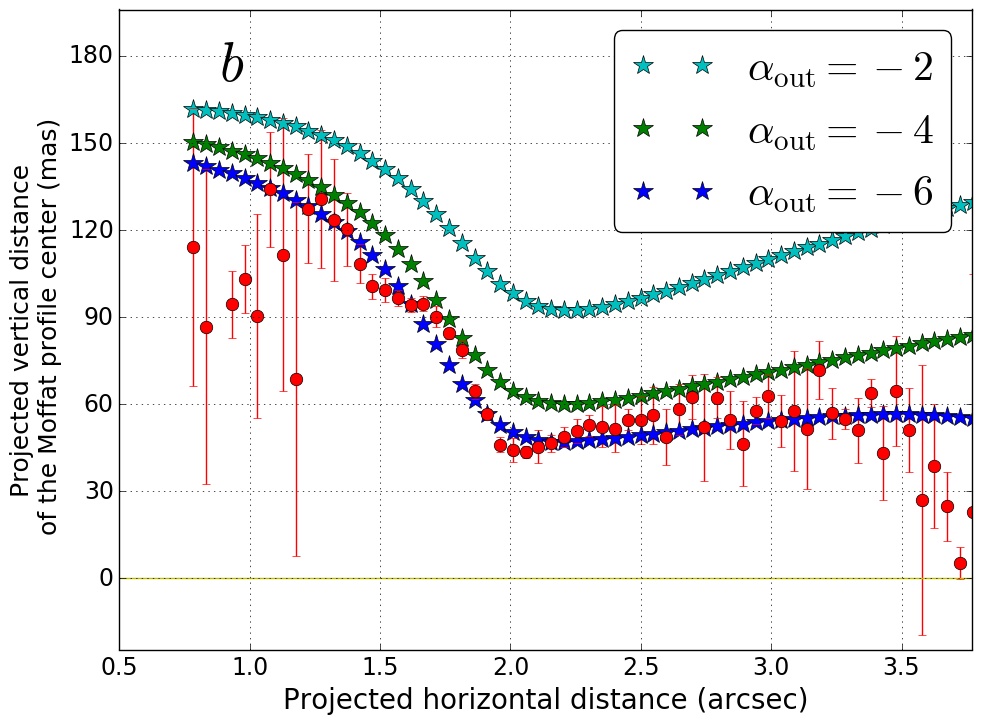}
   \includegraphics[width=8cm]{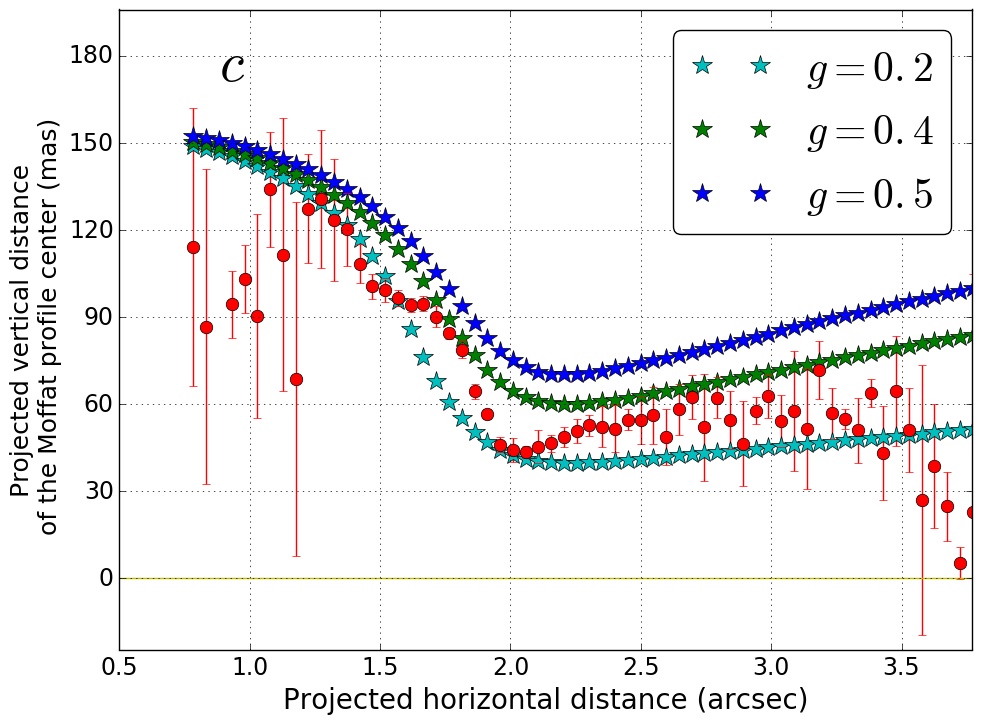}
   \includegraphics[width=8cm]{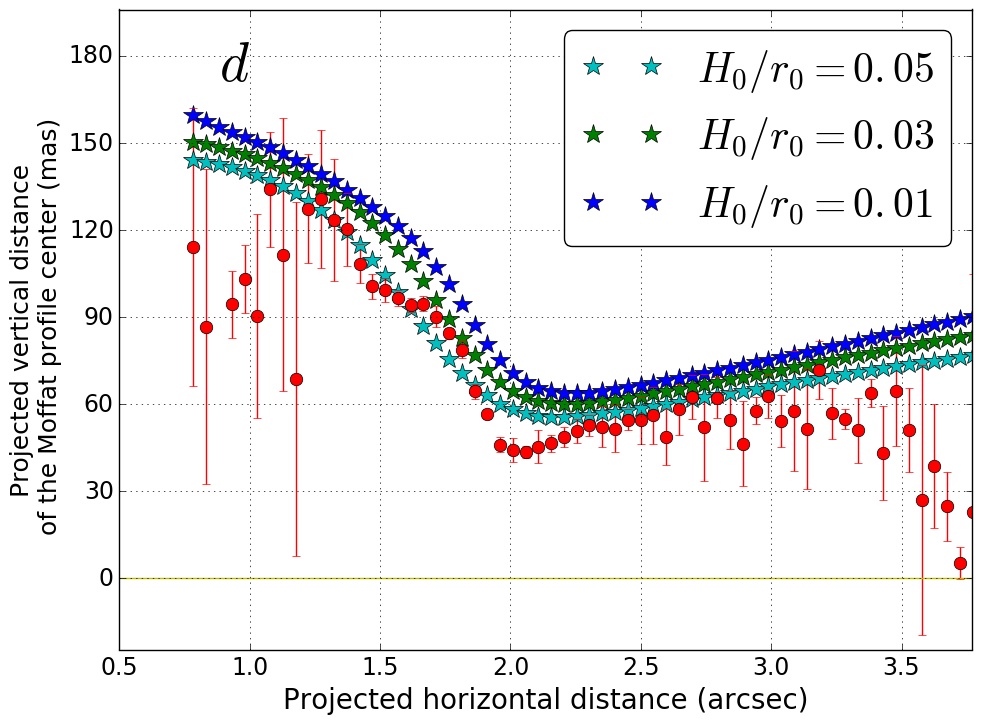}
   \includegraphics[width=8cm]{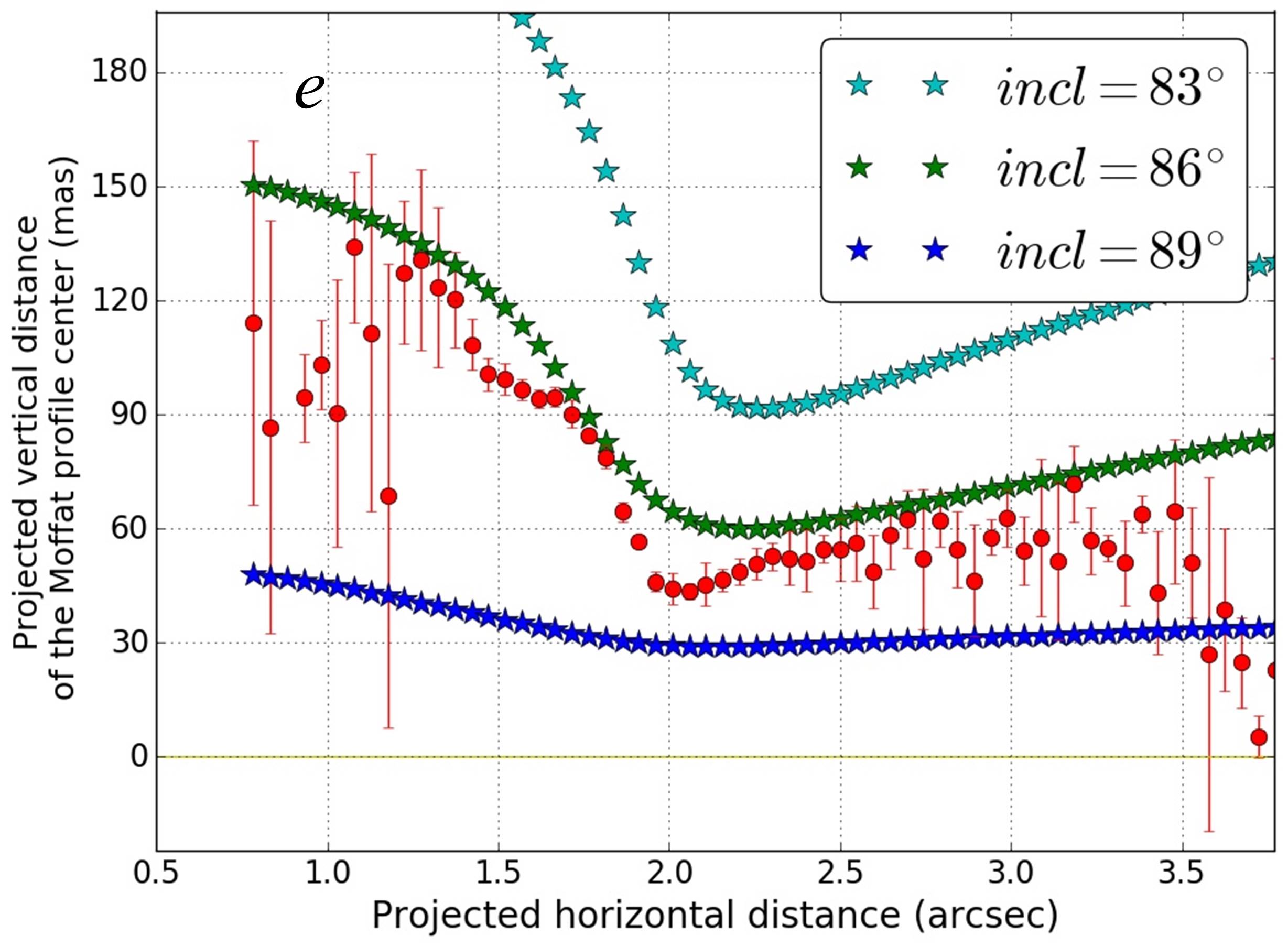}  
   \includegraphics[width=8cm]{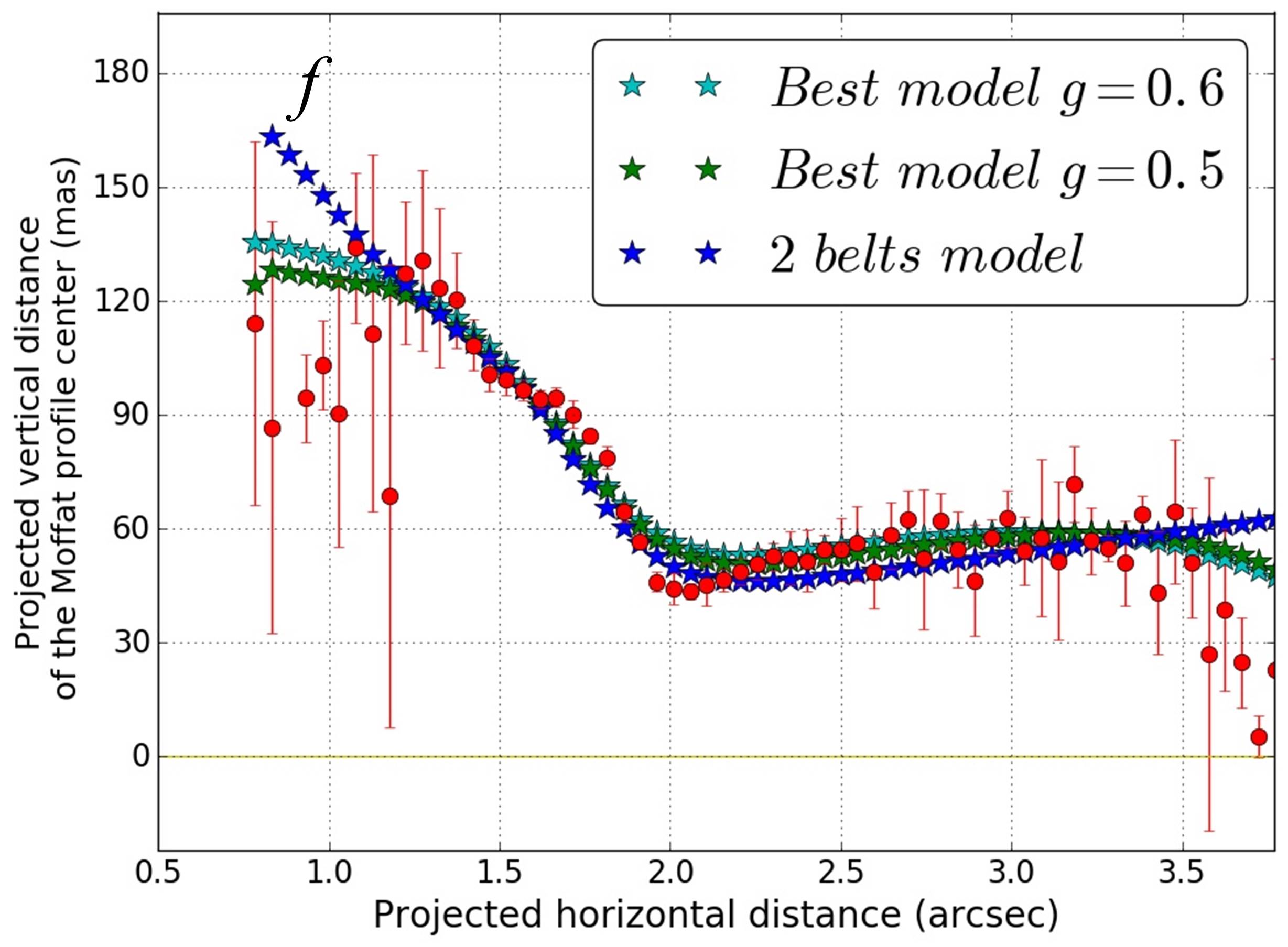}
              \caption{Investigation of the parameter degeneracies between $\alpha_{\rm in}$, $\alpha_{\rm out}$, $g$, $H_0 /R_0,$ and inclination. Red dots mark the location of the spine measured in all 4x4 binned imaging data as a vertical offset of the Moffat profile peak (see Sect.~\ref{s_Spine} for detailed description). {\it a}) Impact of varying $\alpha_{in}$ on the spine location of the model with fixed parameters $g=0.4$, $\alpha_{out}=-4$, $H_0 / R_0 = 0.03$, $incl = 86^{\circ}$. {\it b}) Impact of varying $\alpha_{out}$ on the spine location of the model with fixed parameters $g=0.4$, $\alpha_{in}=6$, $H_0 / R_0 = 0.03$, $incl = 86^{\circ}$. {\it c}) Impact of varying $g$ on the spine location of the model with fixed parameters $\alpha_{in}=6$, $\alpha_{out}=-4$, $H_0 / R_0 = 0.03$, $incl = 86^{\circ}$. {\it d}) Impact of varying $H_0 / R_0$ on the spine elevation of the model with fixed parameters $\alpha_{in}=6$, $\alpha_{out}=-4$, $g=0.4$, $incl = 86^{\circ}$. {\it e}) Impact of varying inclination on the spine location of the model with fixed parameters $\alpha_{in}=6$, $\alpha_{out}=-4$, $g=0.4$, $H_0 / R_0 = 0.03$. {\it f}) Comparison of the two best-fitting one-belt models and the two-belts model with the data. The parameters of the model best tracking the spine are $g=0.6$, $\alpha_{\rm in}=4$, $\alpha_{\rm out}=-7$, $H_0 / R_0 = 0.03$, $incl = 86^{\circ}$. The second best model has $g=0.5$, $\alpha_{\rm in}=3$, $\alpha_{\rm out}=-7$, $H_0 / R_0 = 0.01$ and $incl = 86^{\circ}$.      \label{f_Spine_diagnostics}}
\end{figure*}

Figure~\ref{f_Spine_diagnostics} illustrates the degeneracy between $\alpha_{\rm in}$, $\alpha_{\rm out}$, $g$, $H_0 / R_0$, and disk inclination. 
In each panel, one parameter takes three different values while the others remain constant. 
As can be seen in Fig.~\ref{f_Spine_diagnostics}a, the steeper the dust number density law inside the belt (increasing $\alpha_{\rm in}$), the higher the spine elevation above the midplane inside the parent belt, whereas the steeper fall-off of the grain number density (decreasing $\alpha_{\rm out}$) shifts the spine closer to the midplane (Fig.~\ref{f_Spine_diagnostics}b). On the contrary, the spine is displaced upwards when the asymmetry parameter $g$ increases (Fig.~\ref{f_Spine_diagnostics}c). The impact of the disk aspect ratio $H_0 / R_0$ on the position of the spine is similar to that of the asymmetry parameter - the larger the disk scale height, the larger the spine elevation (Fig.~\ref{f_Spine_diagnostics}d). The vertical offset of the spine from the disk axis can also be affected by the disk inclination. The smaller the inclination, the larger the vertical displacement of the spine and the steeper the curve slope in the inner and outer parts of the disk (Fig.~\ref{f_Spine_diagnostics}e).

Interestingly, we note that the curvature of the spine in the outer region (beyond $r=2''$) changes with decreasing $\alpha_{\rm out}$: the spine is bowed for $\alpha_{\rm out}\leqslant -6$ while it is more like a straight line for larger values. There might be a trend for a bow in the data for $r>3''$ but the S/N of the disk is getting small at these separations. This bow could be partially reproduced by some particular model (Fig. \ref{f_Spine_diagnostics}b).

Figure \ref{f_Spine_diagnostics} clearly demonstrates that the spine is fully determined by a combination of all parameters once the disk radius is fixed. 
Therefore, as an alternative to calculating a 2D $\chi^2$ on scattered light images, which implies a large number of degrees of freedom, we used the spine itself and derived $\chi^2_{\rm spine}$ for angular separations in the range $1-3''$ on the west side of the disk as already done in \citet{Boccaletti2012} for the HD 32297 debris disk.
We define the $\chi^2_{\rm spine}$ according to:
\[\chi^2_{\rm spine} = \frac{1}{N_{\rm data}}\sum_{i=1}^{N_{\rm data}} \frac{\left[ y_{\rm {i \, data}}-y_{\rm {i \, model}}\right]^2 }{ \sigma_{\rm {i \, data}}^2}, \]
where $N_{\rm data}$ is the number of data points between $1''$ and $3''$, $y_{\rm {i \, data}}$ are the vertical offsets of the spine measured from the data with uncertainties $\sigma_{\rm {i \, data}}$ (see Sect. \ref{s_Spine}) and $y_{\rm {i \, model}}$ are the vertical offsets of the model spine. 

The minimal value  $\chi^2_{\rm spine}=2.47$ is obtained for $\alpha_{\rm in}= 4$, $\alpha_{\rm out}=-7$, $g=0.6$, $H_0 / R_0= 0.03$ and $incl=86^\circ$. However, there are other combinations of parameters, in particular the model with $\alpha_{\rm in}= 3$, $\alpha_{\rm out}=-7$, $g=0.5$ with $\chi^2_{\rm spine}=2.53$, which provide a similar good fit to the data. 
As can be seen in Fig.~\ref{f_Spine_diagnostics}f, the spines of these two models show a high degree of similarity with the data. For comparison purposes, the spine of the two-belts model ($\chi^2_{\rm spine}=3.10$), where the inner belt is inclined at $80^{\circ}$ (see Sect.~\ref{s_two_belts}), is also presented (blue asterisks in Fig.~\ref{f_Spine_diagnostics}f). 
Investigating a larger parameter space with the two-belts model should allow to find a better match to the spine curve. 

\begin{figure}  
   \centering
 \includegraphics[width=7.7 cm]{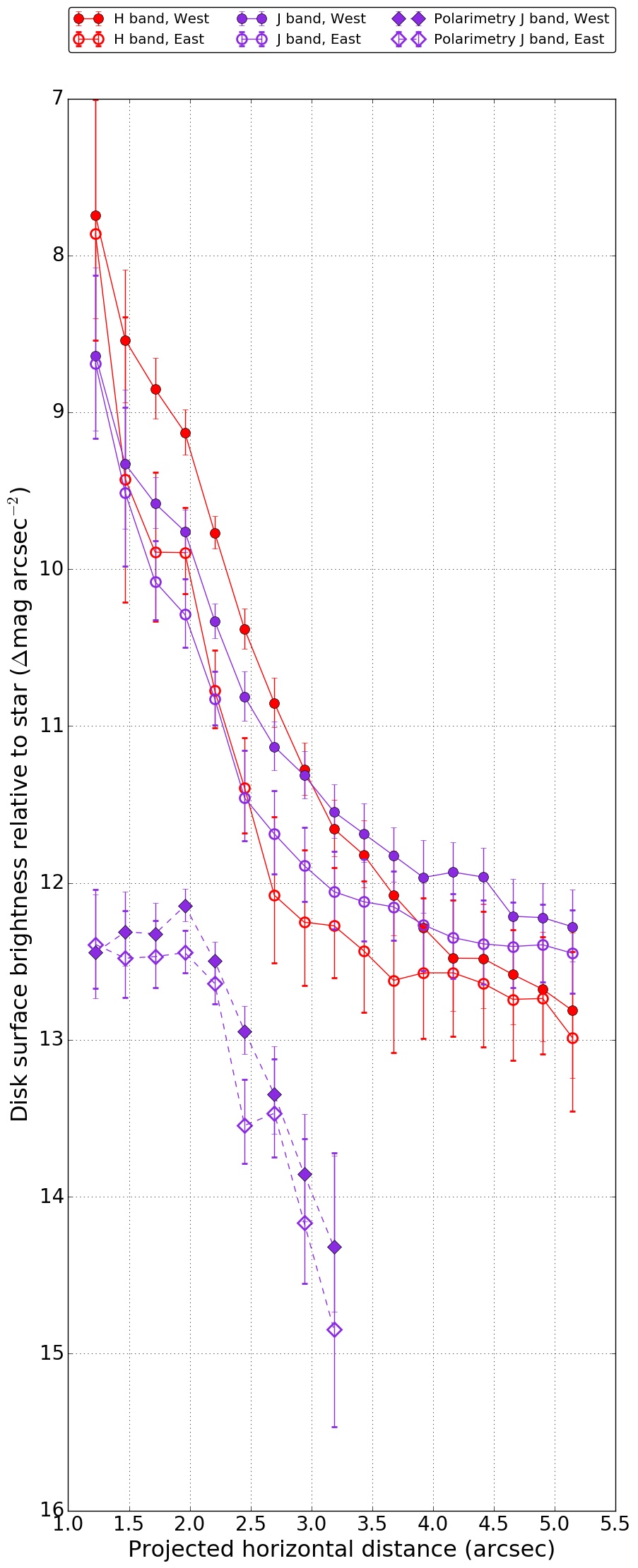}
               \caption{Radial SB profiles of the west and east disk sides in the J and H band. All data points show the disk magnitude measured in square $0.25''\times 0.25''$ ($20 \times 20$ pixels) apertures, centered on the disk major axis, relative to the stellar magnitude in the corresponding filter. Circles indicate the relative magnitude of the total intensity of scattered light (Stokes $I$) and diamonds mark the magnitude of the polarized flux (Stokes $Q_\varphi$). Errors are calculated as a standard deviation of the flux distribution in the concentric annuli.  \label{f_color}}
\end{figure}

\begin{figure}  
   \centering
  \includegraphics[width=7.6 cm]{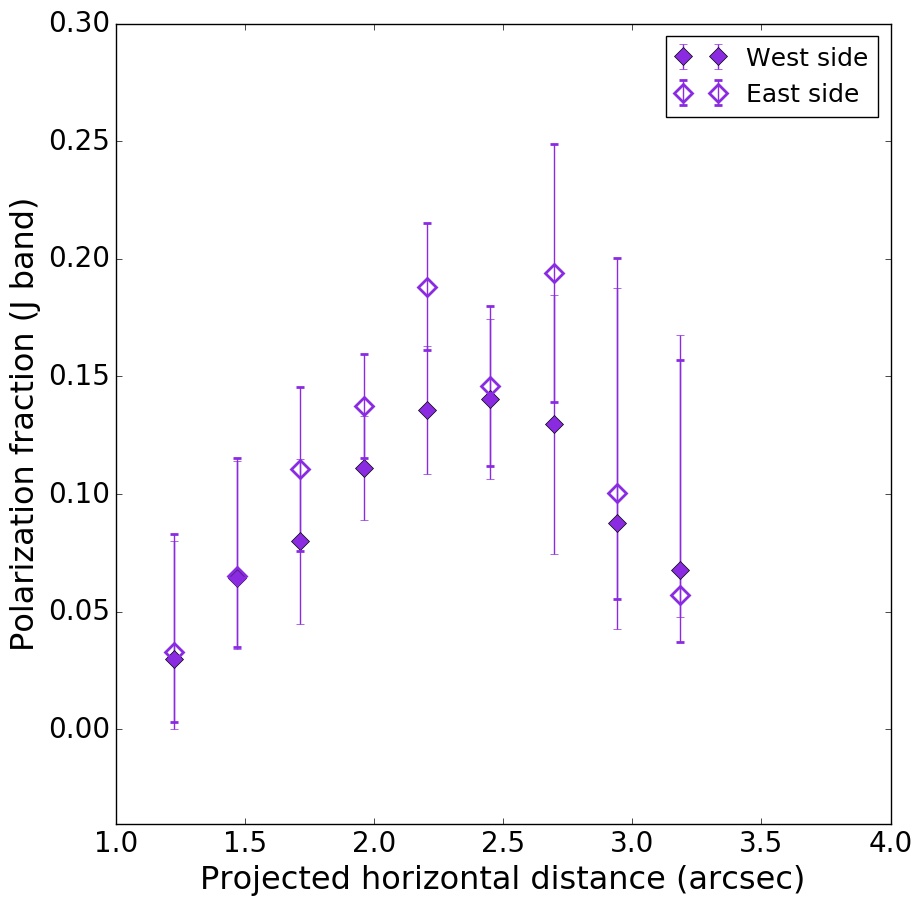}
               \caption{Polarization fraction of the scattered light measured in square $0.25''\times 0.25''$ apertures along the disk axis using total intensity (Stokes $I$) and polarized flux (Stokes $Q_\varphi$) profiles in the broad J band plotted in Fig. \ref{f_color}.        \label{f_Polfraction} }
\end{figure}

\section{Photometry} \label{s_Photometry}
We used the instrumental PSFs for the flux normalisation, in both total intensity and polarimetry. In classical imaging, the two IRDIS channels are summed up for photometric measurements. The stellar flux is measured in a circular aperture centered on the PSF with a radius of $r= 0.6''$ ($r=50$~pixels) after the  subtraction of the background (see Sect. \ref{s_Imaging}). 
Taking the transmission of the neutral density filter ND2.0 and DITs (see Sect. \ref{Observations}) into account, we obtain a stellar count rate of $(7.27\pm 0.80)\cdot10^7$ ADU/s in J band and $(6.52\pm 0.78)\cdot10^7$ ADU/s in H band. 
To measure the disk flux from the total intensity data, we first calculated the correction factor to account for the ADI-induced self-subtraction using the best-fit models obtained for both bands (see Col. 6 in Tables \ref{t_resultsH} and \ref{t_resultsJ}). 
The ratio of the unprocessed and the ADI-processed model images provides the flux reduction factor for each pixel (in the binned images for each binned pixel) inherent to the ADI technique. Hereafter, the disk flux is corrected for the ADI-induced flux reduction. 
The total disk flux is integrated in two rectangular areas containing pixels with ($x,\,y$) coordinates lying within the bounds $1''< |x| < 4''$ and  $-0.14''< y < 0.26''$. We obtain 2900 $\pm$ 300 ADU/s and 4100 $\pm$ 500 ADU/s in J and H, respectively. This yields a brightness ratio between the disk and the star of $(4.0\, \pm \, 0.6)\cdot 10^{-5}$ in the J band and $(6.3 \, \pm \, 1.0)\cdot 10^{-5}$ in the H band for scattered light. 

The polarized flux measured in the same areas in J band is 240 $\pm$ 70 ADU/s. This flux has to be corrected for the reduced instrument throughput in DPI \citep[][SPHERE User manual]{Langlois2010}. Since the flux measurement in DPI  is performed with all instrument components placed in the optical path, we estimate the correction factor by comparing the total stellar flux received in both DPI channels with the stellar flux measured in both CI channels (see Sect.~\ref{Observations}). Calibrating the polarimetric measurement in this way, we assume that the stellar light is unpolarised and that the polarimetric efficiency of the instrument did not significantly change during the observation. We obtain a flux reduction factor in DPI of 3.8 for the left channel and 3.5 for the right channel.
Taking into account the correction factor of 3.5, the actual total polarized flux of the disk amounts to 840 $\pm$ 240 ADU/s. 
With this value, the polarized light contrast of the disk in J band is $(1.2 \pm 0.6)\cdot 10^{-5}$. 
The ratio between the intensity of the polarized light and total intensity of the scattered light is $0.29 \pm 0.09$. 

In the total intensity we measured that the east side flux is $\sim$40\% of the west side flux in both J and H band data (more precise $\sim$38\% and $\sim$44\%, respectively) while in polarimetry $F_{\rm east}= 0.8 F_{\rm west}$. 

\subsection{Disk color}
Total intensity and polarimetric SB profiles are displayed in Fig.~\ref{f_color} for comparison.
Here, the disk flux is measured in square $0.25''\times 0.25''$ (20 $\times$ 20 pixels) apertures centered on the disk axis.
This aperture size was chosen for a better comparison with the radial SB profiles of the HD\,15115 disk in V and H bands presented by \cite{Kalas2007} (their Fig.~3) and in 1.1 $\mu$m band published by \cite{Debes2008} (their Fig.~3). 

For stellocentric distances ranging between $1.5''$ and $3.0''$ we find that the west side is $\sim$ 0.2 - 0.5\,mag$\cdot$arcsec$^{-2}$ brighter than the east side in the J band while this difference is as large as $\sim$ 0.7 - 1\,mag$\cdot$arcsec$^{-2}$  in the H band. 

The steeper fall-off of the SB in the H band ($\propto$~$r^{-3.3}$ in the range $1.2-4''$ on the west side and $\propto$~$r^{-5.9}$ in the range $1.2-3''$ on the east side) compared to the J band ($\propto$~$r^{-2.6}$ in the range $1.2-4''$ on the west side and $\propto$~$r^{-3.5}$ in the range $1.2-3''$ on the east side) produces a change of color at a particular separation to the star. We observe a red or grey color (J-H $\geq$ 0) at short separations, while the color turns to blue (J-H$<$0) at large separations. This color change occurs between 2.5 and $3.5''$ on the west side and between 2 and 3$''$ on the east side. For large separations, J-H can be as large as $-0.7$ mag$\cdot$arcsec$^{-2}$.

It is interesting to note that the polarized light SB profiles of the west and east side show a high degree of similarity: a slow growth between 1.2 and 2$''$ and a fall-off $\propto$~$r^{-3.4}$ for $r>2''$. The SB profiles are truncated because of the low S/N of the polarimetric data at large separations. 

\begin{figure*}  
  \centering
  \includegraphics[width=17cm, trim=-0.9cm 0 0 0cm]{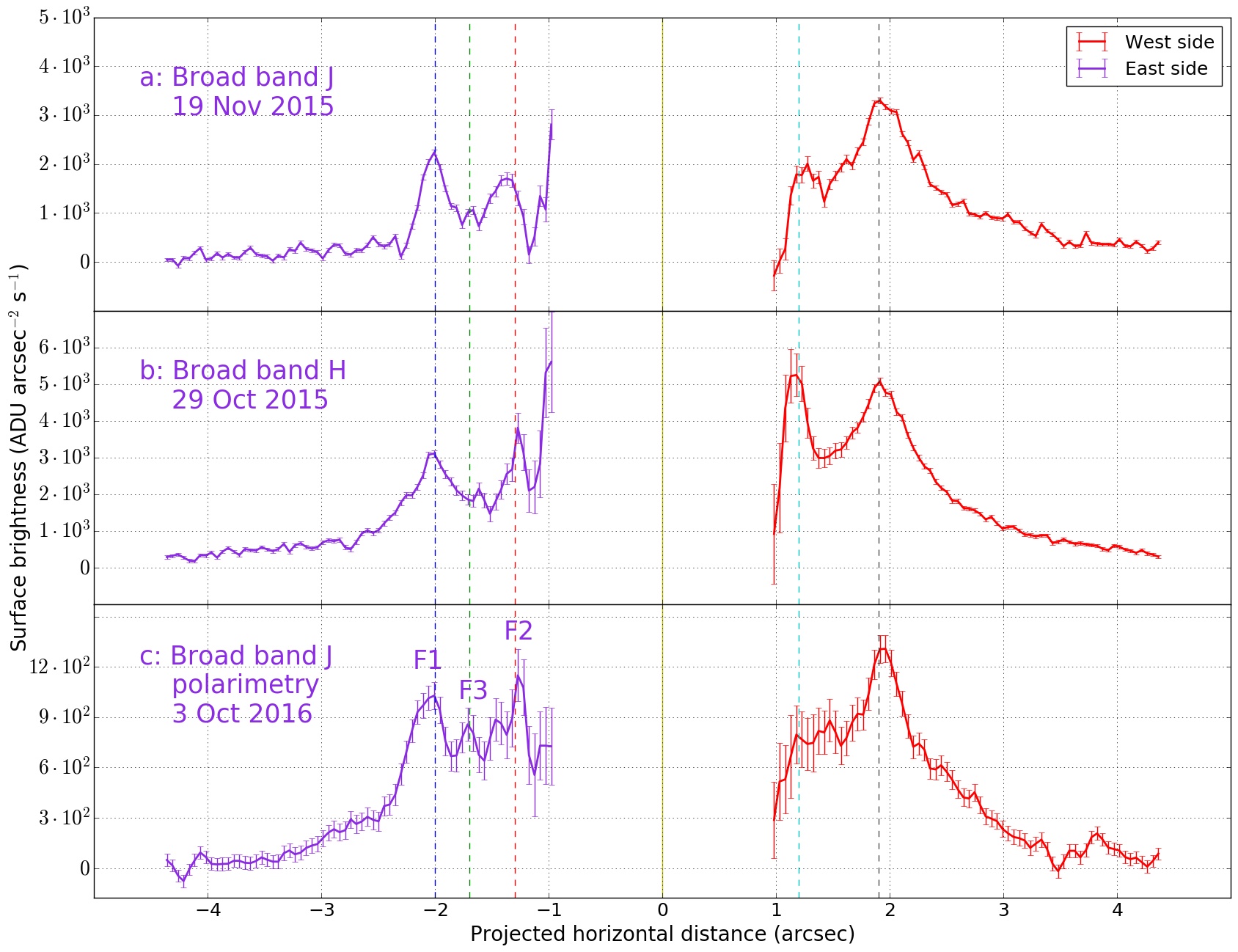} %trim=l b r t
               \caption{Radial SB profiles measured along the disk major axis in $0.049'' \times 0.049''$ square apertures in the cADI image of the J band data (\textit{a}), the cADI image of the H band data (\textit{b}) and in the $Q_\varphi$ image (\textit{c}, J band). All data sets show the SB peaks at $r=2''$ ('F1') and $r\approx1.3''$ ('F2') symmetrically located with respect to the star, except the west disk side in the $Q_\varphi$ image. Images of the disk in Fig.~\ref{f_Pol_J_H_bin4x4} are presented with the same $x$-axis scale for 1 arcsec as in this figure for ease in identification of the SB peak locations highlighted with the vertical dashed lines. \label{f_SB}}
\end{figure*}

\begin{figure*}
  \centering  
 \includegraphics[width=12.2cm, trim=-5.4cm 0 0 0cm]{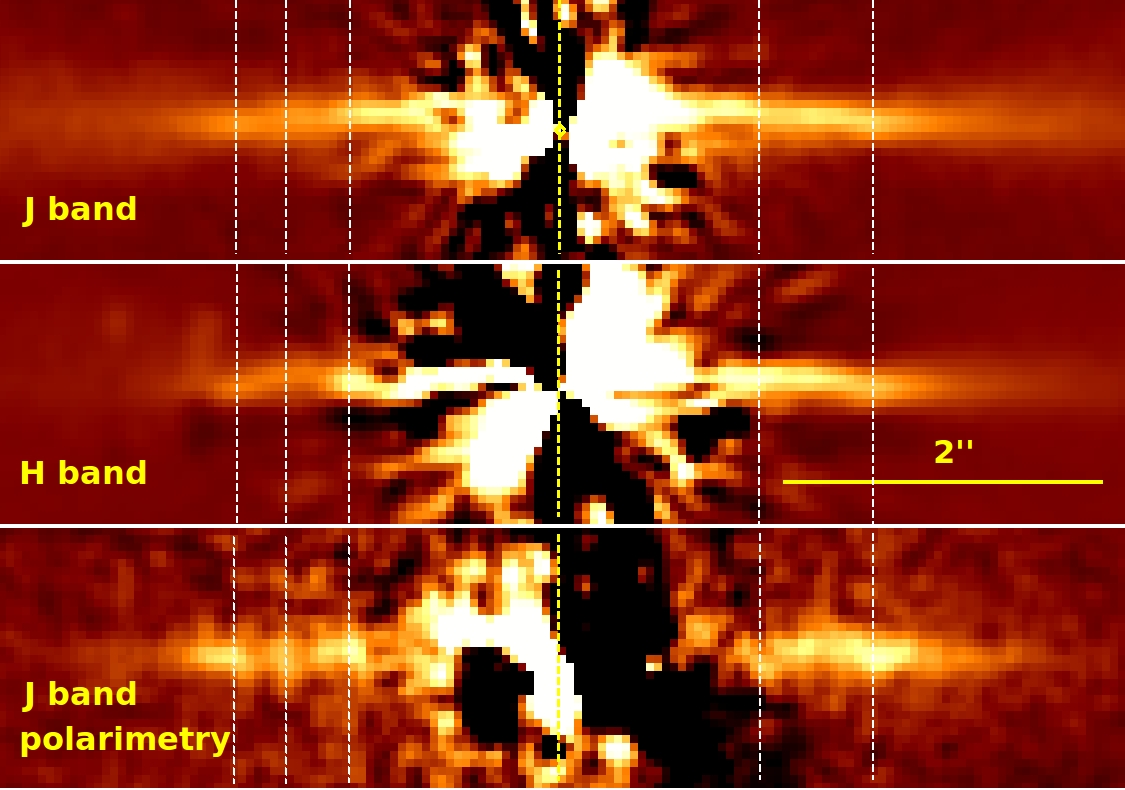}
  \caption{Images of the HD 15115 disk illustrating the position of SB peaks plotted in Fig.~\ref{f_SB}. The original data were $4\times 4$ binned to reduce the noise and smoothed via a Gaussian kernel with $\sigma =$ 2 pixels. The position of the star in each image is marked by a yellow dotted line and corresponds to the position of the star in Fig.~\ref{f_SB} (yellow line). Images are presented in linear scale: J band in [-25, 85], H band in [-120, 400], J band polarimetry in [-10, 34]. \label{f_Pol_J_H_bin4x4}}

\end{figure*}

\subsection{Polarization fraction}
The polarization fraction (PF) of scattered light in the J band is  measured in square $0.25''\times 0.25''$ apertures along the disk axis (Fig. \ref{f_Polfraction}) and is calculated from the ratio between the SB of the polarized and the total intensity data (corrected for the ADI bias). The radial PF profiles of both disk sides show a similar behavior increasing from PF $\approx$ 0.03 at $r=1.25''$ up to $\sim$15-20\% between $r=2''$ and $r=3''$, thus close to the edge of the planetesimal belt. This value is consistent with the degree of polarization in the K band measured by \citet{Tamura2006} along the midplane of the $\beta$ Pictoris debris disk. A similar fractional polarization ($\sim$10-25\%) is obtained by \citet{Asensio-Torres2016} for another edge-on disk HD 32297 in the H band.

At radial separations of $r\gtrsim 3''$, the S/N of our polarimetric data is too low to make a conclusion about the polarization degree of scattered light. It is also not possible to determine whether the PF further increases or decreases, or remains approximately constant with a growing radial distance as was measured for an edge-on debris disk around M star AU Mic \citep{Graham2007}.

\subsection{Surface brightness profiles} \label{s_SBprofile}
The aperture size $0.25''\times 0.25''$ ($20\times20$ pixels) used in the previous sections is too large to see SB variations on smaller scales while the angular resolution provided by IRDIS allows for a more detailed examination of the flux distribution inside $r=2''$.  
In Fig.~\ref{f_SB}, we used a finer sampling of $4 \times 4$ pixels ($0.049''\times 0.049''$) together with a $\sigma =1$ pixel and $\sigma =2$ pixels Gaussian smoothing for classical imaging and polarimetric data respectively.
The errors were computed as the standard deviation of the flux distribution in concentric annuli centered on the star and do not include systematic uncertainty. 

The SB profiles measured in all data sets show a peak at a radial separation of $r\approx 2''$ from the star on both disk sides (denoted as 'F1' in Fig.~\ref{f_SB}c). This peak is clearly attributed to the outer planetesimal ring. All three observations show remarkably consistent results for the radius of this belt: for the east side we measure $R_{\rm east} = 2.00'' \pm 0.05''$  and for the west side $R_{\rm west} = 1.91'' \pm 0.05''$. The small difference in the belt radius between both sides indicates an offset of $\sim 0.045''$ of the belt towards the east with respect to the star.

Further inside the ring, the SB profiles exhibit another peak at $r\approx1.2-1.3''$ also on both disk sides in classical imaging data of J and H bands (denoted as 'F2' in Fig.~\ref{f_SB}c). In polarized intensity data, a SB peak is detected only on the east side at the same radial separation. The location of this peak corresponds to the increased SB at the position indicated by the arrow 'B' in Fig.~\ref{f_QphiUphi}. We observe some kind of correlation between these peaks and think that they could mark the ansae of the inner belt. There is a third SB peak on the east side of the disk ('F3' in Fig.~\ref{f_SB}c) but at a lower significance level. 
 
The increased surface brightness at the locations 'F1', 'F2', and 'F3' is also visible in the disk images in Fig.~\ref{f_Pol_J_H_bin4x4}. For an easier comparison of Figs.~\ref{f_SB} and~\ref{f_Pol_J_H_bin4x4} and better identification of the SB peak locations, the images in Fig.~\ref{f_Pol_J_H_bin4x4} are presented with the same $x$-axis scale for 1 arcsec as in Fig.~\ref{f_SB}. The vertical dotted lines going through the positions of the SB peaks in Fig.~\ref{f_SB} are also shown in Fig.~\ref{f_Pol_J_H_bin4x4}.

\subsection{Phase function} \label{s_Phase_func}
\begin{figure}
   \centering
   \includegraphics[width=8.5cm]{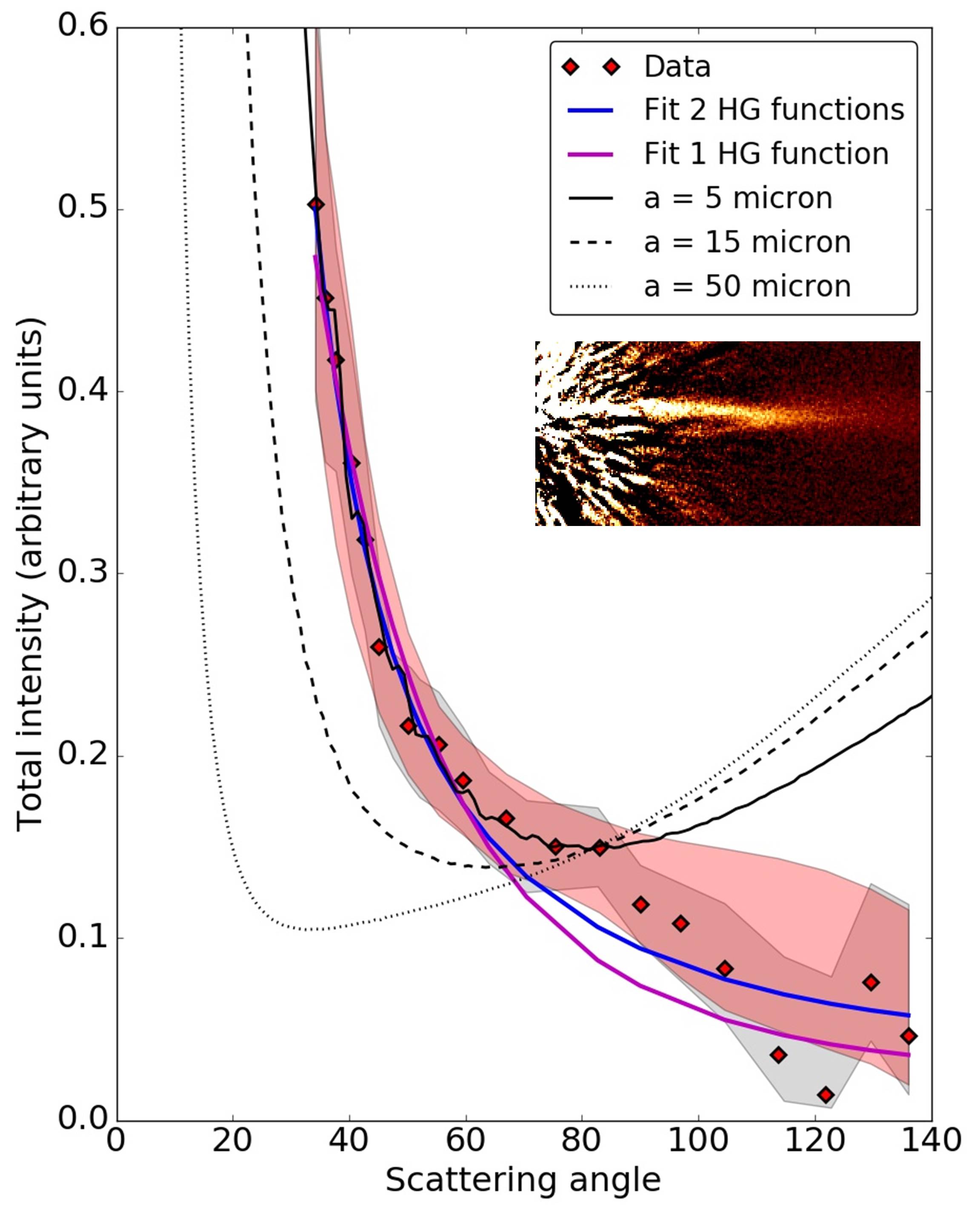}
   \caption{Normalized scattering intensity (red diamonds) measured in circular apertures along the disk spine on the west side shown in the inset image. The blue and purple curves show HG fits to the data points. The black curves show scattering phase functions obtained by \citet{Min2010} for three different sizes of grains $a$ covered with regolith particles. The shaded grey area indicates 1$\sigma$-uncertainties of the measurement. The shaded red area shows the range of values measured from the models. \label{f_Phase_func} }
\end{figure}

In the total intensity data, in particular in the H band dataset, the back side, although faint, is detected. The observed part of the planetesimal ring shown in the inset in Fig.~\ref{f_Phase_func} spans the range of the scattering angles between 35$^\circ$ and 140$^\circ$. Measuring the azimuthal variation of the SB along the rim of the ring, we can constrain the scattering phase function. For the calculation of the scattering angles, we need to make some general assumptions on the grain properties and geometry of the disk. We assume that grains scatter stellar radiation preferentially in the forward direction, this means the northern part of the disk is the near side. We also consider that the circumstellar material is confined to a narrow ring with a radius of $R_0=1.95''$ (94 au). Using the cADI total intensity image in the H band obtained with selected frames (only frames with a seeing condition below 0.9$''$ were used), we performed the aperture photometry of the west disk side in circular apertures with a radius of 6 pixels. The disk area within each aperture was attributed a unique scattering angle which was calculated for the near side of the disk according to: 
\begin{equation}
\theta = \arcsin \frac{\sqrt{x_c^2 + y_c^2}}{R_0}
\end{equation}
and for the back side of the disk
\begin{equation}
\theta_{\rm back} = 180^\circ - \arcsin \frac{\sqrt{x_c^2 + y_c^2}}{R_0},
\end{equation}
where $(x_c, \, y_c)$ are coordinates of the aperture center in the coordinate system shown in Fig.~\ref{f_disks}a. The apertures were centered on the spine of the front and back side of the disk defined as the brightest points of the vertical cross-sections perpendicular to the disk major axis.

To correct for the flux loss caused by the ADI data reduction, the measured flux values were multiplied with a scaling factor derived from the modeling of ADI bias (see Sect.~\ref{s_Photometry}). The errors of this measurement are estimated as one standard deviation of the flux distribution measured in the apertures of the same size which were placed at the same radial distance from the star but at different PAs. 

The obtained intensity profile is shown in Fig.~\ref{f_Phase_func}. The flux values are arbitrarily normalized so that the intensity at 83$^\circ$ equals 0.15. As Fig.~\ref{f_Phase_func} shows, the scattering intensity steeply increases with a decreasing scattering angle for $\theta < 50^{\circ}$ and gives rise to a relatively broad diffraction peak. The measured curve is rather flat between $60^{\circ}$ and $90^{\circ}$, whereas for larger angles corresponding to the fainter back side of the disk ($\theta > 90^{\circ}$) the scattering intensity is decreasing. For comparison, we measured the intensity profiles of several models performing the same aperture photometry. Instead of plotting each model profile individually, we show the range of possible model values with a shaded red area in Fig.~\ref{f_Phase_func}.

We used the $\chi^2$-minimization algorithm to find a HG function (Eq.~\ref{HG}) that matches the intensity profile and fitted the data also with a linear combination of two HG functions following the same approach as in \cite{Milli2017}. According to the $\chi_\nu^2$-criterion, the better fit $f(\theta)$ is given by a linear combination of two HG functions \citep{Engler2017}:
\begin{equation}
f(\theta)= w \, f_{\rm HG \, 1}(\theta, g_1 ) + (1-w) \, f_{\rm HG \, 2}(\theta, g_2),
\end{equation}
with $g_1=0.94 \pm 0.06$, $g_2=0.21 \pm 0.01$ and scaling factor $w = 0.89 \pm 0.10$ (blue curve in Fig.~\ref{f_Phase_func}). For this combination, the reduced $\chi_\nu^2$ metric is 1.22. This is smaller than $\chi_\nu^2$ of 1.43 which is obtained for the best-fit of the HG function with the single asymmetry parameter $g= 0.57 \pm 0.0$ (purple curve). However, the visual comparison of both fits in Fig.~\ref{f_Phase_func} shows that they are nearly identical and thus two additional free parameters of the combined HG functions do not significantly improve the fit. The lower $BIC$ (Sect.~\ref{s_two_belts}) for the HG function with the single $g$ parameter also indicates that this model should be considered as the best-fitting one.

In Fig.~\ref{f_Phase_func}, we also compare the measured intensity profile with three scattering phase functions derived by \citet{Min2010} for dust grains with radii equal to 5, 15 and 50 $\mu$m which are covered by small regolith particles.\footnote{In the original paper by \cite{Min2010}, grain diameters are given in the notation instead of grain radii.} The grains are treated as micro-asteroids reflecting light in the geometric optics regime. Their phase functions are obtained for a narrow grain size distribution and include the diffraction by spherical particles and the reflectance function according to the analytical model for the bidirectional reflectance \citep{Hapke1981}. For a comparison, all three functions in Fig.~\ref{f_Phase_func} are normalized to the data point at $\theta = 83^\circ$. 

At large scattering angles  ($\theta > 90^{\circ}$) the measured phase function of the HD 15115 disk does not follow the phase functions obtained by \citet{Min2010}. When comparing the diffraction parts of the phase functions ($\theta < 80^\circ$), the scattering function for grains with a size of 5 $\mu$m (solid black curve) fits best the measured intensity profile. Assuming that this grain size is the average size $\langle a \rangle$ of the grain size distribution, we derive the minimum grain size $a_{\rm min}= 3$ $\mu$m from the relation $a_{\rm min}= 3/5 \, \langle a \rangle$ which is valid for the dust grains in a collisional cascade equilibrium expected at the location of the parent body ring. This minimum grain size is in agreement with the estimated blowout size for HD 15115 of $\sim$1--3~$\mu$m assuming a stellar luminosity of 3.5$L_\odot$, a mass of 1.3$M_\odot$, and a grain average density in the range from 1 to 3 g cm$^{-3}$ \citep{Rodigas2012, Hahn2010}.

It is interesting to note that the comparison of micro-asteroids scattering functions with the observed phase functions of the debris disks around HR 4796 A and Fomalhaut shows the compliance with the curves for grains with radii of $a=15$~$\mu$m \citep{Milli2017} and $a=50$~$\mu$m \citep{Min2010}, respectively. This could suggest that the grains in the HR 4796 A and Fomalhaut disks are larger on average than the dust grains we observe on the west side of the HD 15115 debris disk. 

\section{Discussion} \label{Discussion}
\subsection{Unpolarized versus polarized light}
As mentioned in Sect. \ref{s_Photometry}, the polarized intensity (Stokes $Q_\phi$) measured for the east side of the disk is a factor of 1.25 lower compared to the west side. In the total intensity data (Stokes $I$) of the J band, we obtain for the east side an intensity which is a factor of 2.67 lower than on the west side. 
The different factors for the total and polarized intensities, when comparing the east side with the west side,
can indicate a difference in the grain number density, grain properties or both. 

In the case of HD 15115, this could be the result of an asymmetric loss of small particles due to the eastward motion of the system through the interstellar medium as suggested by \cite{Debes2009}. 
Such motion could produce a surplus of smaller grains along the line of sight on the west side in front of the planetesimal belt. With polarimetry, we trace predominantly $90^{\circ}$ scattering and therefore the ansae of the belt. The total intensity strongly favours forward scattering from grains in front of the star and the parent belt. This could cause the strong Stokes $I$ signal in the west where the particles produce a lot of ``extra'' scattered intensity $I$ but only a little ``extra'' Stokes $Q_\phi$ signal.

An alternative interpretation of the disk brightness asymmetry was recently proposed by \cite{Lee2016}. They presented a model explaining the ``needle''-like morphologies observed in the edge-on disks by a planet perturbation. Their model requires an eccentric parent belt with a brighter disk side much closer to the star and much shorter in length than the fainter side. Such a model is in contradiction with our data. We observe a symmetric parent belt (see Fig.~\ref{f_SB}) in both J and H bands, in total and in polarized intensity. If the brightness asymmetry of the HD 15115 disk is caused by planets, then the perturbation mechanism should be more complex than the model suggested by \cite{Lee2016}.

\subsection{Color}
The photometric analysis (Figs. \ref{f_color} and \ref{f_SB}) suggests that inside the planetesimal belt ($r\lesssim 2''$) the disk color is red and becomes bluer with increasing distance from the star ($r>2''$). 
A similar color behavior has previously been observed for the HD 15115 disk by \citet{Kalas2007} and \citet{Debes2008}.
They argued that the change in the dust scattering efficiency beyond $r\approx2''$ could be due to the absorption features of dust material at 1.6 $\mu$m and inside $r\approx2''$ due to absorption of olivine at $\sim$1 $\mu$m, for instance. Another possibility to explain the complex color of the disk would be the existence of dust populations with different average grain sizes varying with distance from the star. At the radial position of the ``birth'' ring we would expect to find dust grains larger on average than the blowout size $a_{\rm blow}$. Beyond the ``birth'' ring small dust particles might dominate the scattering cross section due to the radiation blow-out of the smallest dust grains \citep[][and references therein]{Stark2014, Krivov2010}. Assuming that the collisional cascade power law for the grain size distribution $dn(a)\propto a^{-3.5}\,da$ is valid at the location of the parent body ring, we would expect that this distribution becomes steeper, and therefore the disk color bluer, with distance from the star. 

\begin{figure}  
 \includegraphics[width=8cm]{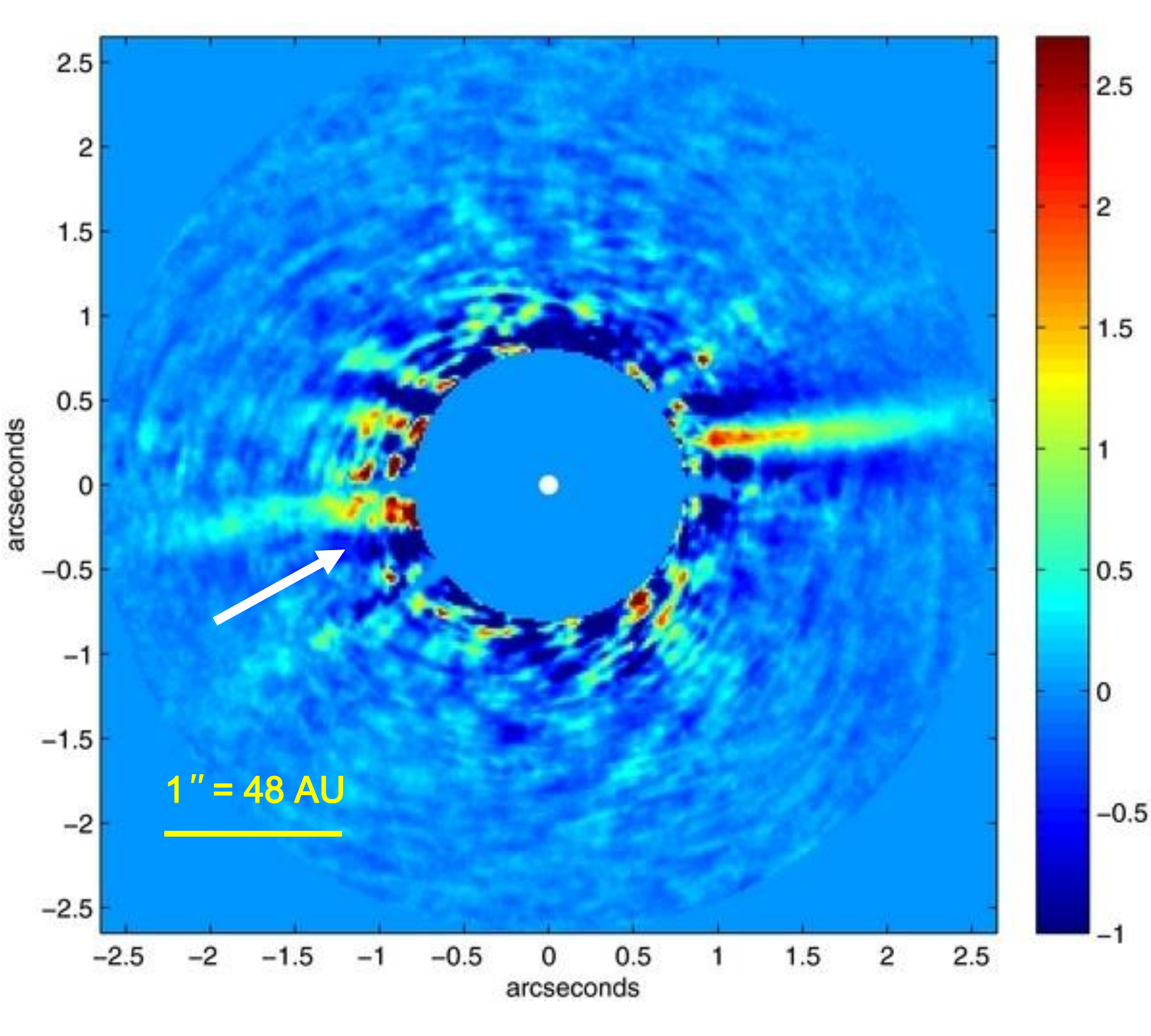} 
               \caption{K$_S$ band image of the HD\,15115 debris disk as published by \cite{Rodigas2012}. \label{f_Rodigas}}
\end{figure}
\begin{figure}  
   \centering
   \includegraphics[width=7.9cm]{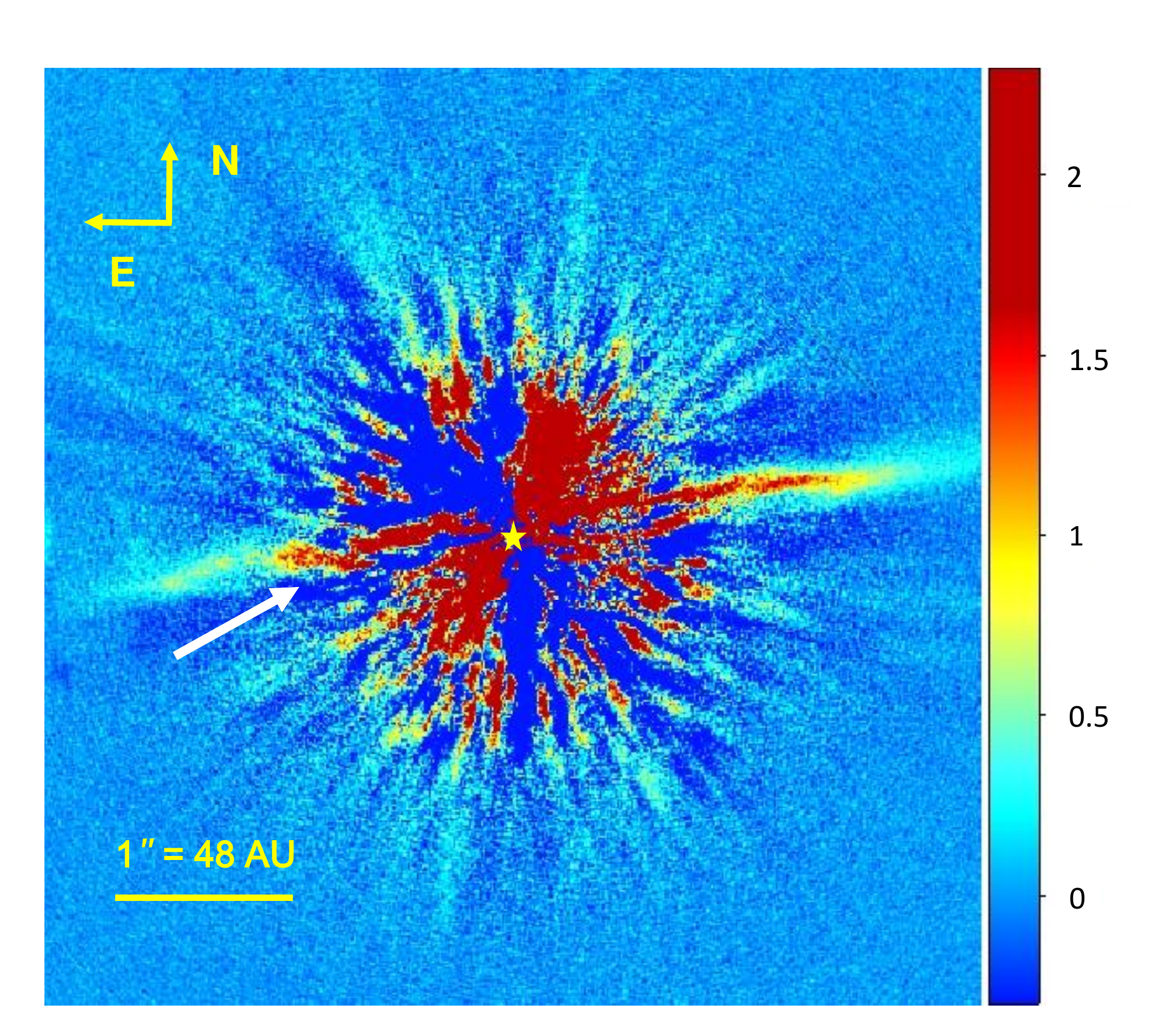}
   \caption{H band image of the HD\,15115 debris disk (cADI, this paper).  \label{f_Ellipse}}
\end{figure}

\subsection{Evidence for multiple belts?}
Our images show an axisymmetrical disk in agreement with the observations of \cite{Kalas2007} and \cite{Rodigas2012}. \cite{Mazoyer2014} have shown that in H and K$_S$ filters, the disk has a symmetric ring shape with a radius of $R_0\approx 2''$ and an inclination of $\sim$86$^{\circ}$ well matching the geometrical parameters which we derive for the ring observed in the broad bands H and J (Fig.~\ref{f_imaging}). This ring could be an outer planetesimal belt of a multiple ring system.

The disk morphology seen in our polarimetric data (Sect. \ref{Morphology}, Fig. \ref{f_disks}) suggests the presence of a second inner planetesimal ring which could have a slightly different PA and inclination compared to the outer belt. The structural clumps that we see in the $Q_\varphi$ image (Fig. \ref{f_QphiUphi}b) and ripples of the SB profile along the disk axis, in particular on the east side of the disk (Fig. \ref{f_SB}) are similar to the ripples in the SB profile of the polarized light detected by \citet{Tamura2006} in the $\beta$ Pic debris disk. The 2 $\mu$m polarimetry of the $\beta$ Pic disk has revealed many details in the inner part of the disk, including the previously observed warp. In analysing the disk structure in the region between 50 and 121 au, \citet{Tamura2006} concluded that the ripples of the polarimetric SB profile may indicate a multiple ring system. We think that we see the same effect in the SB profile of the polarized light in Fig.~\ref{f_SB}c. It is obvious that the outer SB peaks (denoted as 'F1' in Fig.~\ref{f_SB}c) mark the location of the outer ring ($R\approx 2''$) of the HD\,15115 disk. Therefore the other SB peaks, which we measure further inside the disk, may also indicate inner rings. 

The modeling of the total intensity and polarized intensity data in both J and H bands does not rule out the two-belts model. According to the $\chi^2$ criterion for the goodness-of-fit, which we have calculated for more than 2000 models, the two-belts model provides a good fit to the total intensity data being always among the 20 best models. For the H band dataset, it is even the best model when fitting both disk sides simultaneously (see Table \ref{t_resultsH}). 

In Appendix \ref{appendix_SED}, we show that the two-belts model is also consistent with the spectroscopic and photometric data of HD 15115, if the modified blackbody function \citep{Backman1993} is used to fit the disk SED.

There are some indications from previous studies that support the idea of HD\,15115 as a multiple belt system:
\begin{itemize}
\item As already mentioned in Sect.~\ref{s_Introduction}, \cite{Moor2011} and \cite{Chen2014} claimed that the disk SED can be better fitted with a combination of two blackbody models which implies the presence 
of two planetesimal belts. However, both authors found that an inner belt has a small radius of 4 or 5 au. Therefore this belt cannot be the same inner belt suggested by us.
\item \cite{Debes2008} and \cite{Rodigas2012} measured an increasing PA 
of the eastern disk side with decreasing distance
to the star for $r< 1.6''$ (see their Figs. 2a and 4a, respectively) which is in 
contradiction to a bow-like disk shape reported by \cite{Rodigas2012} and \cite{Schneider2014}.
Both authors suggested a nearly edge-on disk model with a high HG 
scattering asymmetry parameter to explain the "bowing" of the brighter (northern) disk side.
In accordance with this model, the PA of the east disk side should decrease 
following the course of the disk spine when approaching the star. 
This means that the vertical offset of the spine from the disk axis towards the north should increase for smaller separations, as is expected for a single belt with a high inclination and a high asymmetry parameter for forward scattering (see Fig. 12 in \cite{Schneider2014} for an illustration).  
The measured vertical offset of the spine is decreasing in observations at 1.1 $\mu$m \citep[][their Fig. 2a]{Debes2008} and in K$_S$ band for $r<1.6''$ \citep[][their Fig. 4a]{Rodigas2012}. These measurements could be affected by the lower spatial resolution of the PISCES and NICMOS data and by their very special structure of the stellar halo with more low spatial frequencies, which could mimic a disk structure. However, our data also show a small decrease of the spine offset between $r=1.5''$ and $r\approx$1.25$''$ on the east side in the total intensity images of both J and H bands (see Fig.~\ref{f_Spine}).\\
Furthermore, \cite{Rodigas2012} measured a clear break in the radial SB power law at $r \approx 1.4''$ on the east side of the disk: from a power law index of -1.1 for $r > 1.4''$ to a power law index of -5.27 for $1.2'' < r < 1.4''$ (see their Fig.~5a). The power law break of the radial SB profile is often connected with the location of the planetesimal belt. The rapid increase of the SB could be due to stellar residuals or due to an additional source of emission inside $r \approx 1.5''$ which could be a planetesimal ring. In Fig.~\ref{f_Rodigas}, we reproduce Fig.~1a from \cite{Rodigas2012}
presenting a K$_S$ band image of the debris disk; it shows the increasing 
surface brightness below the disk axis close to the coronagraph only on the east 
side (indicated with a white arrow in Fig.~\ref{f_Rodigas}). We see a similar feature
in our H band data at the same position (see Fig.~\ref{f_Ellipse}). It might be responsible for the small dip in the spine curve (see Fig.~\ref{f_Spine}) mentioned above. 

\item The inner belt with a slightly different PA from the outer belt can easily explain the bifurcation of the east disk side observed by \cite{Schneider2014} and \cite{Debes2008} (see their Fig. 1; feature A is explained as a contamination from the PSF subtraction but it roughly coincides with the bifurcated structure). The radial blowout of the smallest dust grains from two planetesimal belts at different PAs can create the same morphology.
\end{itemize}

It is also possible that the SB peak, which we observe to the east at $r \sim$1.3$''$ from the star (indicated with a white arrow in Fig.~\ref{f_QphiUphi}b and with the arrow 'B' in Fig.~\ref{f_disks}b) and attribute to the east edge of the inner belt, is actually associated with some clumpy feature of the outer planetesimal belt. Assuming that this dust clump moves on the Keplerian orbit, we could observe its orbital movement when its deprojected radial position changes by $\sim$0.2$''$ (clump width).
Such a position shift on a circular orbit of the radius $r=2''$ for a 1.3$M_\odot$ star would be observable in about 18.5 years.  

\section{Summary} \label{s_Summary}
In this paper, we present classical imaging and polarimetric observations of the HD 15115 debris disk known as the ``blue needle'' with SPHERE-IRDIS in the broad bands J and H. The total intensity images (Stokes $I$, Fig. \ref{f_imaging}) in both bands were extracted with ADI using cADI, KLIP, and LOCI processing algorithms. Images of the polarized scattered light (Stokes $ Q_\varphi$, Fig. \ref{f_QphiUphi}) in the broad band J were obtained using the double difference method. 

Our results from the analysis of the disk geometry and photometry can be summarized as follows:
\begin{itemize}
\item The overall disk structure observed in both bands is an axisymmetric, nearly edge-on, debris ring consistent with the detection by \cite{Mazoyer2014}. The radius of the disk towards the east is $R_{\rm east} = 2.00'' \pm 0.05''$ ($\sim$96 au) and towards the west it is $R_{\rm west} = 1.91'' \pm 0.05''$ ($\sim$92 au). By fitting an ellipse to the spine curve we derive a disk inclination of $85.8^{\circ} \pm 0.7^{\circ}$. 

\item The PA of the disk major axis measured in the total intensity and polarized intensity data is $\theta_{\rm disk}=278.9^{\circ} \pm 0.1^{\circ}$.

\item The disk photometry and radial SB profiles along the disk major axis show the east-west asymmetry. In two rectangular areas bounded by the coordinates $1''< |x| < 4''$ and  $-0.14''< y < 0.26''$ in the total intensity data of J and H bands, we measure $\sim$2.5 times less flux on the east side compared to the west side. In the same areas of the polarimetric data taken in the J band, this difference is smaller and the polarized flux on the east side amounts to $\sim$80\% of the flux measured on the west side (Sect. \ref{s_Photometry}). This might indicate a different average grain size in the disk to the east compared to the disk to the west due to the loss of small dust grains towards the west. 

\item Based on the forward modeling results, we measure a change in the disk color between 2$''$ and 3.5$''$ from red (for $r<2''$) to predominantly blue (for $r<3.5''$). This color difference may indicate different grain size distributions at the location of the parent body belt and in the outskirts where the smallest grains are swept out of the disk.

\item The image of the polarized light from the disk exhibits some SB peaks which can also be seen in the radial SB profile along the disk axis (Fig. \ref{f_SB}c). Similar SB peaks can be recovered in the total intensity data (Figs. \ref{f_SB}a and \ref{f_SB}b). They may be indicative for the presence of a second (inner) planetesimal ring. Based on the J band polarimetry, we estimate that such a ring would have a radius of between $1.3''$ and $1.5''$ and a slightly different PA from the outer ring. 

\item We have performed extensive model calculations to investigate the compliance of the two-belts model with the total intensity and polarized intensity data. According to the $\chi^2$ criteria, both the one-belt and the two-belts models provide similar goodness of the fit. Therefore, the existence of the inner ring cannot be rule out. Taking into account the relatively low S/N of our data, no robust conclusion can be made.
\end{itemize}

To prove the presence of the second belt in the inner regions, it would be valuable to re-observe the HD 15115 disk with ZIMPOL/SPHERE polarimetry. An observation at shorter wavelengths would provide a required high resolution of $\sim$25-30 mas when observing in Very Broad Band (VBB, $\lambda_c=735$~nm) or I band ($\lambda_c=790$~nm). The Strehl ratio is higher in the I band but the count rate is $\sim$40\% lower compared to the VBB. Under good observing conditions (seeing $\leqslant 0.7$, coherence time $ \geqslant 3.5$ ms) 
% and wind speed $\sim 2-3$ m/s) 
a high contrast of $10^{-6}$ can be achieved. Polarimetry as a differential technique has also a decisive advantage over ADI in that it does not suffer from the disk self-subtraction, which could be crucial if we want to study the close-in disk structure.

\begin{acknowledgements}
We would like to thank the referee for many thoughtful comments which helped to improve this paper. This work is supported by the Swiss National Science Foundation
through the grant number 200020 - 162630. We acknowledge the financial support from the Programme National de Plan\'{e}tologie (PNP) of CNRS-INSU co-funded by the CNES, and by the Agence Nationale de la Recherche (ANR-14-CE33-0018). J.\,O. acknowledges financial support from the ICM (Iniciativa Cient\'ifica Milenio) via the N\'ucleo Milenio de Formaci\'on Planetaria grant, from the Universidad de Valpara\'iso, and from Fondecyt (grant 1180395).  J.\,M. acknowledges support from the NASA through the NASA Hubble Fellowship grant HST-HF2-51414.001 awarded by the Space Telescope Science Institute, which is operated by the Association of Universities for Research in Astronomy, Inc., for NASA, under contract NAS5-26555.
\end{acknowledgements}

\bibliographystyle{aa} % style aa.bst 

\bibliography{reference.bib} 

\newpage 

\appendix

\section{Signal-to-noise ratio}
\begin{figure*} 
   \centering
 \includegraphics[width=17.5cm]{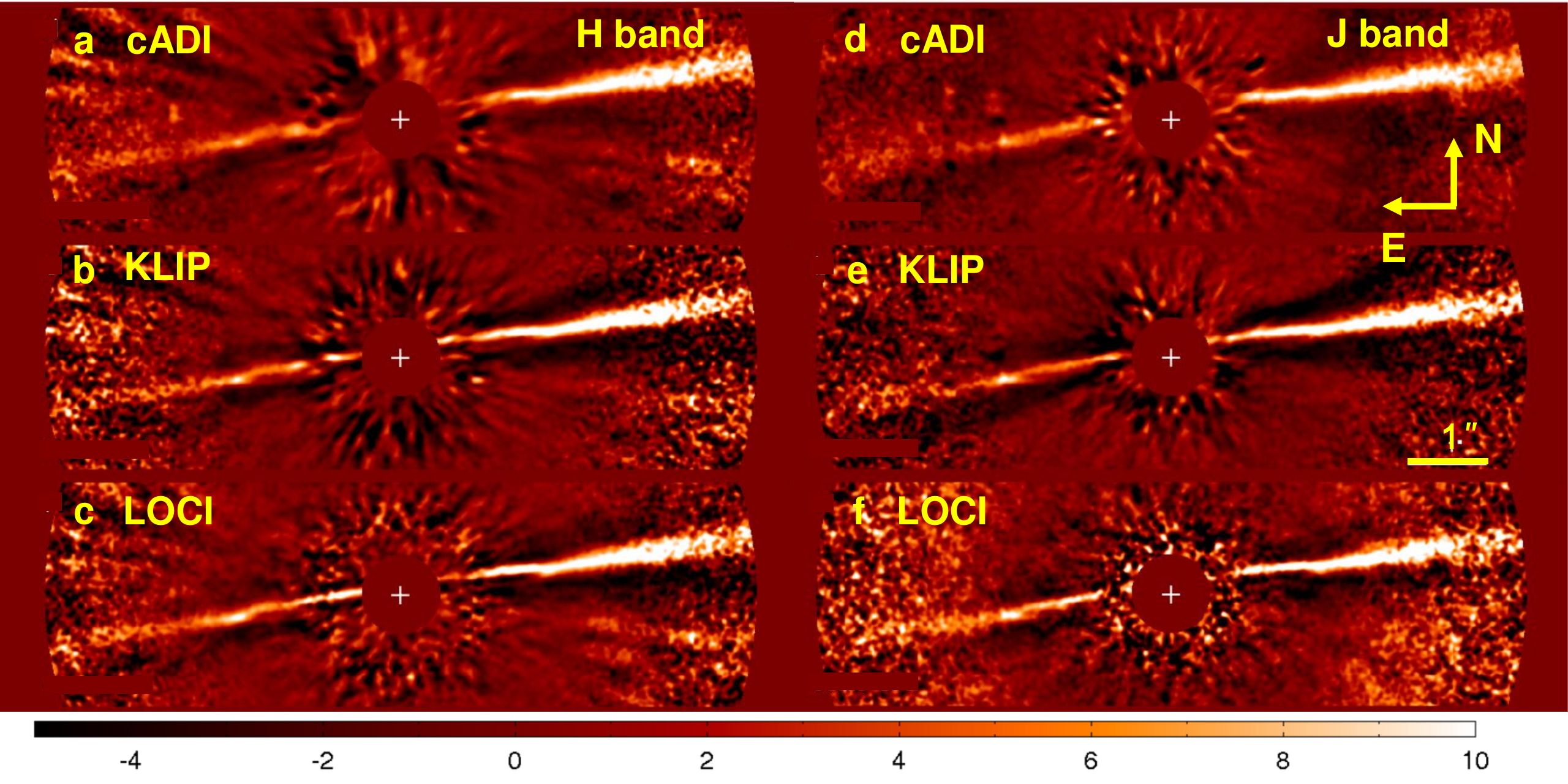}   
  \caption{S/N map for different data-reduction techniques in the H band ({\it left column}) and in the J band ({\it right column}). The number of modes in KLIP reductions is 10. The star position is marked with a white cross. 
 \label{f_imaging_snr}    }
 \end{figure*}

Figure \ref{f_imaging_snr} demonstrates the S/N of the total intensity images in J and H bands.
 
\section{Position angle of the disk}
\begin{figure}
   \centering
   \includegraphics[width=8.5cm]{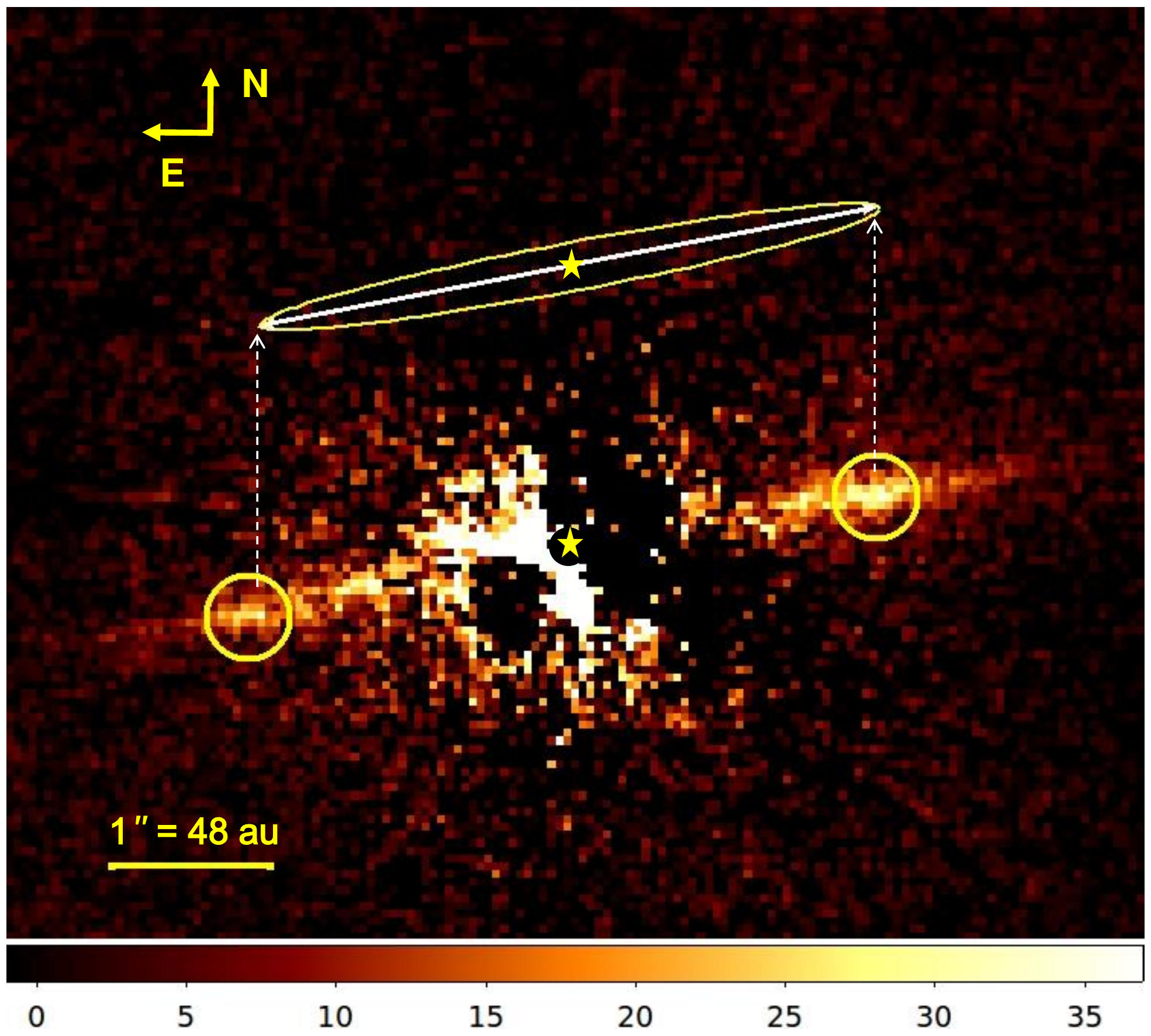}
    \caption{$Q_\varphi$ image illustrating the method to determine the PA and size of the outer disk. The image was $4\times 4$ binned and smoothed via a Gaussian kernel with $\sigma =$ 1 pixel. The brightest pixels of the regions inside the yellow circles indicate the location of the planetesimal belt which is schematically drawn as a yellow ellipse. For the clarity, the ellipse was shifted upwards (white dotted lines show the parallel shift of the brightest pixels). The position of star is marked by a yellow asterisk. The white line represents the disk major axis. The color bar shows the counts per binned pixel. \label{PA} }
 \end{figure}

See Fig.~\ref{PA}.
 
\section{Consistency of the two-belts configuration with the disk SED} \label{appendix_SED}
To check the consistency of the two-belts model with the available spectroscopic and photometric data of the HD 15115 disk, we fitted the disk SED with a single temperature blackbody (BB) emission $B_{\nu}(\lambda, T_{\rm dust})$ and with the modified Planck function \citep[MBB,][]{Backman1993}:
\begin{equation*}
B_{\nu, \,{\rm mod}}(\lambda,\, T_{\rm dust}) = B_{\nu}(\lambda, T_{\rm dust}) \times (H(\lambda_0 - \lambda)+H(\lambda - \lambda_0)(\lambda_0/\lambda)^{\beta}),
\end{equation*}
where $B_{\nu}(\lambda, T_{\rm dust})$ is the Planck function defined by the wavelength $\lambda$ and the dust temperature $T_{\rm dust}$, $H$ is the Heaviside step function, $\lambda_0$ is the characteristical wavelength, above which the absorption efficiency of dust grains scales like $(\lambda_0/\lambda)^{\beta}$ with $\beta$ being the opacity index. The free parameters $\lambda_0$ and $\beta$ allow to account for the steeper falloff of the debris disk spectra at longer wavelengths than the Planck function and to have a higher dust temperature than the blackbody temperature. The free parameter of the MBB model are the dust temperature $T_{\rm dust}$, the opacity index $\beta$, the characteristic wavelength $\lambda_0$ and the scaling factor. In the BB model, the belt radius and thus the dust temperature are fixed. The only free parameter of this model is the scaling factor for the Planck function. We adopted a belt radius of 2$''$ (96 au) with $T_{\rm dust}=39$ K for the one-belt model, and additionally a radius of the second belt of 1.3$''$ (61 au) with $T_{\rm dust\, inn}=48$ K for the two-belts model. To fit the thermal emission from two planetesimal belts, we used a linear combination of two BB or two MBB functions, assuming the same optical properties and average grain size in both belts. 

The disk SED is constructed using the sampled 5--35 $\mu$m \textit{Spitzer}/IRS spectrum \citep{Moor2011} and fluxes compiled from the literature (Table ~\ref{t_SED}). The stellar photosphere emission of a star is calculated with  $L_\ast = 3.5L_\odot$ and $T_{\rm eff} = 6780 K$ \citep{Moor2011}.

\begin{figure}
   \centering
   \includegraphics[width=9.3cm]{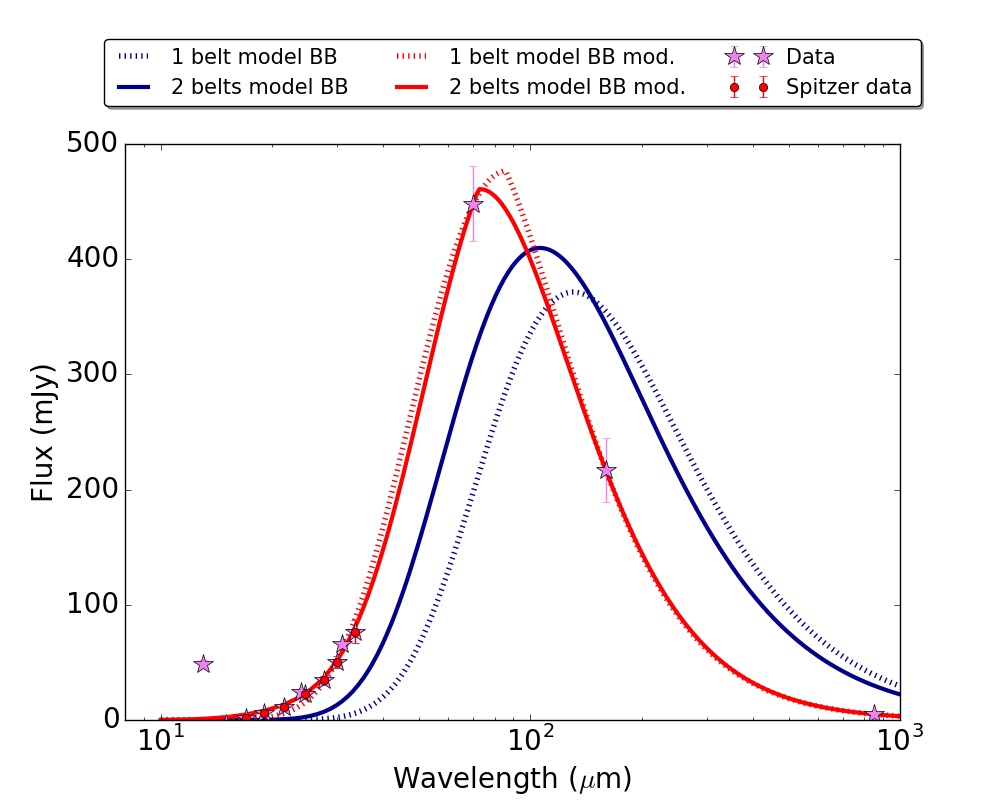}
    \caption{Disk SED and the best fits for the BB and MBB models with one and two belts. Red asterisks show the sampled 5--35 $\mu$m \textit{Spitzer}/IRS spectrum. Violet asterisks represent the photometric data specified in Table~\ref{t_SED}. \label{f_SED} }
\end{figure}

The derived best fits for each model are plotted in Fig.~\ref{f_SED}. The single and double temperature BB models (blue lines) cannot reproduce the data well. The contribution of the lower temperature component in the two-belts BB model to the derived best fit (blue solid line) is negligible. Contrary to that, both MBB models, with one and two belts (red lines), respectively, fit the data very well except for one data point at 13 $\mu$m \citep{Chen2014} which also disagrees with the \textit{Spitzer}/IRS spectrum (AOR 10885632) retrieved from the archive. 

\begin{table}  
     \caption[]{Photometric measurements of HD 15115. \label{t_SED}}
	     \centering
         \begin{tabular}{ccccc}
            \hline
            \hline
            \noalign{\smallskip}
       Wavelength & Flux density & Error & Instrument & Ref.\\  
       ($\mu$m) & (mJy) & (mJy) & & \\   
           \hline
            \noalign{\smallskip}
       13  & 170.2 & 2.9 & \textit{Spitzer}/IRS  & 1 \\[5pt]
	   24  & 58.3  & 2.3 & \textit{Spitzer}/MIPS & 2 \\[5pt]
	   31  & 85.6  & 2.3 & \textit{Spitzer}/IRS  & 1\\[5pt]
       70  & 451.9  & 32.6 & \textit{Spitzer}/MIPS & 2 \\[5pt]
       160 & 217.3  & 27.8 & \textit{Spitzer}/MIPS & 2 \\[5pt]
       850 & 4.9  & 1.6 & JCMT/SCUBA & 3 \\[5pt]
	   \noalign{\smallskip}
           \hline
            \hline
            \noalign{\smallskip}
         \end{tabular}
\begin{flushleft}
{\bf References.} (1) \cite{Chen2014}; (2) \cite{Moor2011}; (3) \cite{Williams2006}. 
\end{flushleft}
\end{table}

\end{document}